\documentclass[english,aps]{revtex4-1}
\usepackage{mathptmx}
\usepackage[T1]{fontenc}
\usepackage[latin9]{inputenc}
\usepackage{color}
\usepackage{babel}
\usepackage{array}
\usepackage{booktabs}
\usepackage{multirow}
\usepackage{amstext}
\usepackage{amssymb}
\usepackage{graphicx}
\usepackage{esint}
\usepackage[unicode=true,pdfusetitle,
 bookmarks=true,bookmarksnumbered=false,bookmarksopen=false,
 breaklinks=false,pdfborder={0 0 0},pdfborderstyle={},backref=false,colorlinks=true]
 {hyperref}
\hypersetup{
  linkcolor = blue,urlcolor  = blue,citecolor = blue,anchorcolor = blue,pdfauthor={Thomas Engels}}

\makeatletter

\providecommand{\tabularnewline}{\\}

\@ifundefined{textcolor}{}
{%
 \definecolor{BLACK}{gray}{0}
 \definecolor{WHITE}{gray}{1}
 \definecolor{RED}{rgb}{1,0,0}
 \definecolor{GREEN}{rgb}{0,1,0}
 \definecolor{BLUE}{rgb}{0,0,1}
 \definecolor{CYAN}{cmyk}{1,0,0,0}
 \definecolor{MAGENTA}{cmyk}{0,1,0,0}
 \definecolor{YELLOW}{cmyk}{0,0,1,0}
}

\makeatother

\begin{document}

\title{The impact of turbulence on flying insects \\ in tethered and free
flight: \\high-resolution numerical experiments}

\author{Thomas Engels}
\email{thomas.engels@ens.fr}

\affiliation{LMD-CNRS, École Normale Supérieure and PSL, Paris, 24 rue Lhomond,
75231 Paris Cedex 05, France}

\author{Dmitry Kolomenskiy}

\affiliation{Japan Agency for Marine-Earth Science and Technology (JAMSTEC), 3173-25
Showa-machi, Kanazawa-ku, Yokohama Kanagawa 236-0001, Japan. }

\author{Kai Schneider}

\affiliation{Aix--Marseille Université, CNRS, Centrale Marseille, I2M UMR 7373,
Marseille, 39 rue Joliot-Curie, 13451 Marseille Cedex 20 France }

\author{Marie Farge}

\affiliation{LMD-CNRS, Ecole Normale Supérieure and PSL, Paris, 24 rue Lhomond,
75231 Paris Cedex 05, France}

\author{Fritz-Olaf Lehmann}

\affiliation{Department of Animal Physiology, Universität Rostock, Rostock, Albert-Einstein-Str.
3, 18059 Rostock, Germany}

\author{Jörn Sesterhenn}

\affiliation{ISTA, Technische Universität Berlin, Berlin, Müller-Breslau-Strasse
12, 10623 Berlin, Germany }

\date{16 January 2019}
\begin{abstract}
Flapping insects are remarkably agile fliers, adapted to a highly
turbulent environment. We present a series of high resolution numerical
simulations of a bumblebee interacting with turbulent inflow. We consider
both tethered and free flight, the latter with all six degrees of
freedom coupled to the Navier--Stokes equations. To this end we vary
the characteristics of the turbulent inflow, either changing the turbulence
intensity or the spectral distribution of turbulent kinetic energy.
Active control is excluded in order to quantify the passive response
real animals exhibit during their reaction time delay, before the
wing beat can be adapted. Modifying the turbulence intensity shows
no significant impact on the cycle-averaged aerodynamical forces,
moments and power, compared to laminar inflow conditions. The fluctuations
of aerodynamic observables, however, significantly grow with increasing
turbulence intensity. Changing the integral scale of turbulent perturbations,
while keeping the turbulence intensity fixed, shows that the fluctuation
level of forces and moments is significantly reduced if the integral
scale is smaller than the wing length. Our study shows that the scale-dependent
energy distribution in the surrounding turbulent flow is a relevant
factor conditioning how flying insects control their body orientation.
\end{abstract}
\maketitle

\section{Introduction}

Insect are fast and agile fliers, which stabilize their body posture
during flight under a vast variety of environmental conditions \citep{Liu2017,Chapman2011}.
While flight in static air requires little steering and corrective
changes in aerodynamic force production, flight in turbulent air is
challenged by unexpected changes in flow conditions at the body and
wings. Little is known about the impact of turbulence on the aerodynamic
performance and energetic cost of flight in insects. In this work,
we study how different kinds of perturbations affect flapping fliers
in free flight. 

In contrast to laminar flows, turbulent flows are dominated by nonlinear
interactions and, as a result, excite fluctuations on a wide range
of scales. After averaging the flow in either ensemble, time or space,
we identify different length scales characteristic for the turbulent
regime. From large to small, these classical scales are: (i) the integral
scale $\Lambda$ where, on average, the velocity the strongest, and
where therefore energy transport is most active, (ii) the Taylor microscale
$\lambda$ where, on average, the velocity gradients are most intense,
(iii) the Kolomogorov scale $\eta$ below which, on average, the flow
fluctuations are damped by the fluid viscosity \citep{Pope2001}.

In nature, unsteady turbulent flow conditions significantly vary depending
on the terrain and weather conditions. The \textquotedblleft flight
boundary layer\textquotedblright , characterized by conditions favorable
for insect flight, can span for up to 1500 meters above the ground
level in warm weather \citep{Chapman2010}. Activity such as long-distance
migration is typical of high altitudes while foraging, for example,
mainly takes place in the vegetation layer up to several meters above
the ground. This diversity of flow conditions, besides variation in
the temperature, density and mean wind speed, exposes flying insects
to a variety of turbulent flows, ranging from those dominated by wakes
and canopy-layer turbulence at low altitude \citep{Finnigan2000,Raupach1988,Raupach1981},
to the atmospheric turbulence determined by weather and wind systems
at high altitude \citep{Wyngaard1992,Wyngaard2010}.

Until now, studies have focused on selected model organisms such as
hummingbirds \citep{Ortega-Jimenez2014a}, moths \citep{Hedrick2006,Ortega-Jimenez2013},
bumblebees \citep{Ravi2013,Ravi2016}, etc., subject to archetypal
air flow conditions such as von Kármán wakes \citep{Ortega-Jimenez2013,Ravi2013,Ravi2016}
or grid turbulence \citep{Combes2009}. In the hawkmoths \emph{Manduca
sexta}, for example, yaw and roll oscillations of the animal body
are synchronized with the vortex shedding frequencies in the wake
behind a large cylinder \citep{Ortega-Jimenez2013}. Vortex shedding
in von Kármán wakes, however, differs from turbulence since, at moderate
Reynolds numbers, vortices are shed periodically in time and the flow
has strong spatial correlations. Few numerical \citep{Engels2015}
and experimental \citep{Combes2009,Crall2016} studies addressed flapping
flight in turbulent flow and estimated flow conditions at body and
wings. In heavy turbulence, for example, bumblebees are highly prone
to changes in roll stability and crash if roll velocity exceeds a
maximum value \citep{Combes2009}.

High maneuverability in flight is likely key in coping with turbulence,
at the cost of low stability \citep{Hedrick2009,Hedrick2006}. Insects
that stabilize their body posture during aerial perturbations thus
require fast feedback responses. These responses may rely on passive
and active changes of wing- and body kinematics. Both mechanisms might
help to mitigate aerial perturbations. While passive changes of wing
kinematics result from the interplay between wing material properties
and inertial/aerodynamic forces and thus elastic wing deformation
\citep{Lehmann2011,Nakata2012,Nakata2012a,Ennos1989,Rueppell1989,Combes2003}
active control is imposed by the sensomotor system of the animal and
thus changes in flight muscle activation \citep{Dudley2002}. A complex
passive mechanism has been reported for the fruit fly \emph{Drosophila}
\citep{Beatus2015}. The latter study suggests that wing rotation
about the wing\textquoteright s longitudinal axis \citep{Lehmann2011}
behaves like a system composed of a damped torsion spring. The animal
might control wing rotation by actively changing stiffness and damping
coefficients of this spring, as well as the resting feathering angle.
In this case, fluid--structure interaction results from a combination
of passive changes \emph{via} spring deformation and active changes
\emph{via} modifications of the spring\textquoteright s elastic property.

Besides passive changes, insects also possess a large variety of active
control mechanisms for body stabilization and flight heading control.
Studies on flight control highlighted several unique mechanisms of
wing motion modulation in insects (see \citep{Dickinson2016} for
a recent review). These mechanisms include changes in stroke amplitude,
stroke frequency, stroke plane, the wing's angle of attack and timing
of wing rotation at the end of each half stroke \citep{Combes2009,Fry2003}.
Freely flying bumblebees, for example, stabilize body roll by changes
of the relative difference between left and right stroke amplitude
\citep{Crall2016}. Insects also actively change body shape that modifies
their inertia tensor during flight. Fruit flies \citep{Berthe2015},
hawkmoths \citep{Noda2014} and chestnut tiger butterflies \citep{Yokoyama2013},
for example, change and stabilize their flight heading by changing
the angle between thorax and abdomen.

Previous studies considered the body roll axis of an animal to be
most susceptible for aerodynamic perturbations, owing to its small
moment of inertia compared to yaw and pitch. Flying through turbulence
thus produces largest fluctuations about the roll axis in insects
\citep{Ravi2013}. To minimize these changes, some insects laterally
extend their hind legs that increases the roll moment of inertia \citep{Combes2009}.
Although this behavior has been found in orchid bees, smaller insects
such as the fruit fly benefit only little from this mechanism owing
to their small legs \citep{Berthe2015}. Since the hind legs of orchid
bees are untypically large compared to other insect species, it is
less likely that the latter mechanism represents a common mechanism
for roll control in insect flight \citep{Berthe2015}. A most significant
mechanism to cope with air turbulence is aerodynamic damping, resulting
from the flapping wing motion. It is termed flapping-counter-torque
\citep{Hedrick2009,Hedrick2006,Cheng2010} and primarily acts in the
direction perpendicular to the stroke plane. In a horizontal stroke
plane, roll damping only occurs if left and right wings flap at different
angle of attack \citep{Faruque2010}. In an inclined stroke plane,
the moment vector is deflected from the vertical and contributes to
roll dynamics, even during symmetrical motion of both wings. The concept
of flapping-counter-torque in insects was extended to damping coefficients
for all six degrees of freedom of body motion \citep{Cheng2011}.

To understand body posture control of insects flying in turbulent
air, we here present a numerical study. Our study models and compares
flight of both a tethered and freely flying insect (bumblebee). We
consider different turbulent flows and vary their turbulence intensity
as well as their characteristic length scales, \emph{e.g.}, the integral
scale. Under free flight conditions, the model insect is allowed to
translate along and rotate about all three body axes, in response
to aerodynamic, inertial and gravitational forces, and moments, respectively.
However, we exclude any active control in this work.

Our previous study \citep{Engels2015} showed that in tethered flight
even strong inflow turbulence has little effect on mean force production
and moments, and thus on aerodynamic mechanisms. Building on this
finding, we here explore the effects of turbulent length scales on
a freely flying insect model and demonstrate the effect of turbulence
on body posture in free flight. The approach allows body motion but
ignores any \emph{passive} deformation, of both body and wing, and
also \emph{active} steering. Our study investigates if and how the
scale-dependent energy distribution is relevant for body orientation
control in flying insect.

The complicated time-dependent geometry and the resulting complex
flow topology challenge numerical simulations of insect flight. There
are two major numerical approaches for this problem. (i) Overset grids
\citep{Liu1998,Liu2009,Sun2002}, which allow strong refinement near
surfaces, but considering inflow turbulence is practically excluded
because of difficulties in parallelization and hence limited resolution.
(ii) Immersed Boundary Methods (IBM) which disconnect the flapping
motion from the grid and thus simplifies the discretization. For flapping
flight, finite volume \citep{Medina2015,Kim2001} or lattice-Boltzmann
type simulations \citep{Suzuki2011,Hirohashi2017} are successful
numerical methods combined with IBM. Here, we use the volume penalization
method combined with a Fourier pseudospectral solver \citep{Engels2015a}.
This numerical method is characterized by the absence of numerical
dissipation, its high efficiency on massively parallel computers thanks
to the optimized implementation of FFTs \citep{Pekurovsky2012} and
the possibility to impose turbulent inflow.

The remainder of the manuscript is organized as follows: the computational
setup is illustrated in section \ref{subsec:Numerical-wind-tunnel}
and the characteristics of inflow turbulence are described in section
\ref{subsec:Inflow-turbulence}. Section \ref{subsec:Bumblebee-model}
presents the bumblebee model and section \ref{subsec:Governing-equations-and}
recalls the governing equations and briefly outlines the numerical
method. The results and discussion section \ref{sec:Results-and-discussion}
presents first tethered flight simulations and then different free
flight cases. Finally, conclusions are drawn in section \ref{sec:Conclusions-and-perspectives}
and some possible directions for future work are proposed.

\section{Flow Configuration and Numerical Method}

\begin{figure}[h]
\noindent \begin{centering}
\includegraphics[width=0.8\textwidth]{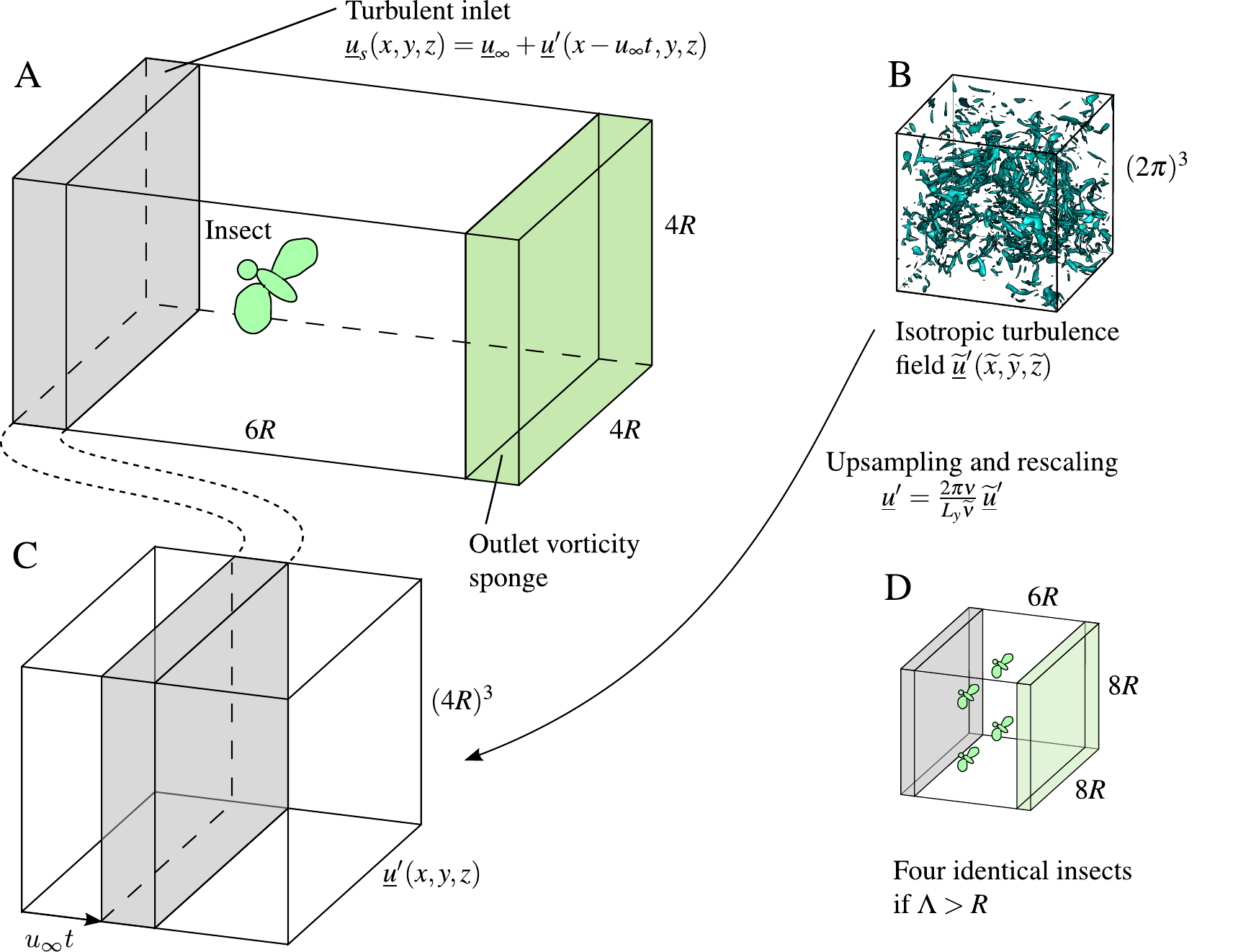}
\par\end{centering}
\caption{Setup used in present work.\label{fig:Setup-detailled-description}
(A) Numerical wind tunnel, with tethered or freely flying insect.
Turbulent inflow is imposed in the upstream gray area, a vorticity
sponge in the downstream green area damps vortices and thus minimizes
their upstream influence. The turbulent inlet imposes a slice of an
isotropic turbulence field (B), which has been pre-computed in a separate
simulation. The turbulence field has been upsampled (C) to match the
resolution of the numerical wind tunnel and rescaled preserving dynamic
similarity. The gray slice in (C) moves through the periodic field
$\underline{u}'(x,y,z)$ at constant speed $u_{\infty}$. In some
simulations with larger integral scale $\Lambda$, four identical
insects are computed (D) in one simulation, as explained in section
\ref{subsec:Inflow-turbulence}.}
\end{figure}

\subsection{Numerical wind tunnel\label{subsec:Numerical-wind-tunnel}}

We illustrate the computational set up and the flow configuration
in Fig. \ref{fig:Setup-detailled-description}. Simulations are performed
in a $6R\times4R\times4R$ large virtual wind tunnel (Fig. \ref{fig:Setup-detailled-description}A),
where $R$ is the wing length of the insect (see section \ref{subsec:Bumblebee-model}).
We initially place the insect at $\underline{x}_{\mathrm{cntr}}=(2R,\,2R,\,2R)^{T}$
and either allow it to move freely as dictated by the fluid forces
or tether it to that position. The resolution in space is $1152\times768\times768$
equidistant grid points, thus the lattice spacing
is $\Delta=5.2\cdot10^{-3}R$. The mean flow velocity is set to $\underline{u}_{\infty}=(1.246\,Rf,\,0,\,0)^{T}$,
where $f$ is the wing beat frequency. It compensates for the cruising
speed of the insect in laminar flow. We initialize the simulation
with unperturbed, laminar flow, $\underline{u}(\underline{x},t=0)=\underline{u}_{\infty}$.
At the outlet, a vorticity sponge \citep{Engels2014} minimizes the
upstream influence of the periodicity of the computational domain.
In the inlet region, which covers the first 48 grid points, the velocity
is set to $\underline{u}_{s}=\underline{u}_{\infty}+\underline{u}'$
, where $\underline{u}'$ are velocity fluctuations obtained from
a precomputed, homogeneous isotropic turbulence (HIT) velocity field
(Fig. \ref{fig:Setup-detailled-description}B). The properties of
this field are discussed in section \ref{subsec:Inflow-turbulence}.

We rescale the HIT velocity field to insect dimensions preserving
dynamic similarity, as HIT simulations are typically performed in
a dimensionless manner. The field is then upsampled using zero-padding
in Fourier space to match the resolution of the numerical wind tunnel
(Fig. \ref{fig:Setup-detailled-description}C). Note that the resolution
requirement for the bumblebee is larger than for the HIT simulations
in all considered cases, as required by the detailed geometry of the
bumblebee. In cases with larger integral scale, we compute four identical
bumblebees in one simulation with doubled lateral domain size and
the same resolution (Fig. \ref{fig:Setup-detailled-description}D),
for reasons explained below.

\begin{figure}[h]
\noindent \begin{centering}
\includegraphics[width=0.8\textwidth]{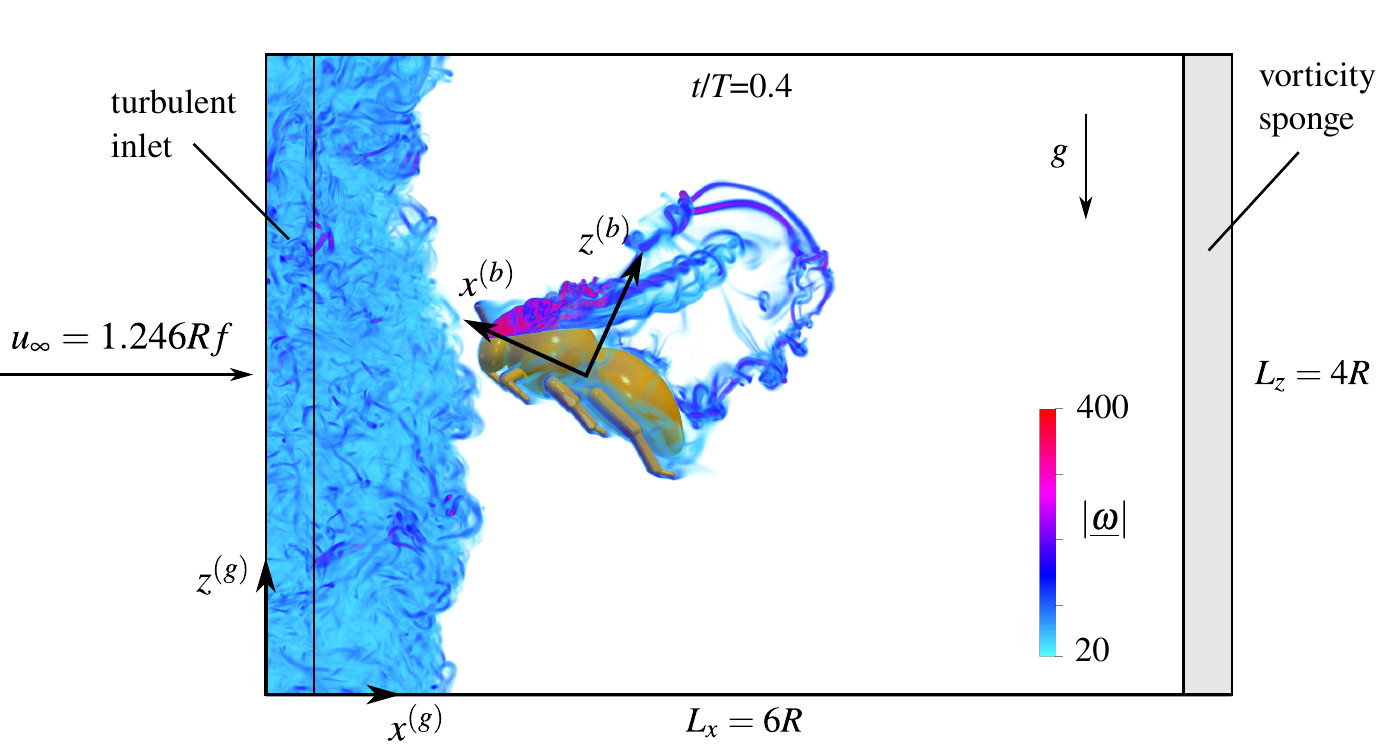}
\par\end{centering}
\caption{Snapshot of a simulation. The virtual bumblebee is tethered in the
virtual wind tunnel, the mean flow $u_{\infty}$ in $x^{(g)}$ direction
compensates for the cruising speed of the insect. Reference frames
shown are the global (g) and body-fixed (b). The flow field is visualized
by the vorticity magnitude at an early instant ($t=0.4T$) before
the laminar/turbulent interface reaches the insect. The parameters
are $I=0.99$ and $\Lambda=0.77R$. \label{fig:snapshot-vorticity}}
\end{figure}

Fig. \ref{fig:snapshot-vorticity} shows an example computation. Inside
the inlet layer, the HIT field is frozen, \emph{i.e}., not dynamically
evolving. Further downstream the turbulent flow evolves dynamically,
and decays similar to what is observed in grid turbulence. The imposed
constant mean flow $u_{\infty}$ transports the turbulent/laminar
interface, as illustrated in Fig. \ref{fig:snapshot-vorticity}. It
reaches the insect's head at $t/T=0.95$ and its tail at $t/T=1.95$.
Thus all wing beats after the second one take place in turbulence
and are used to compute the statistics. After $t/T=3.21$, the periodic
HIT field repeats, owing to the spatial periodicity of the precomputed
field. For each statistical state of inflow turbulence, we compute
a number of realizations $N_{R}$ to be able to perform ensemble averaging.
All simulations are identical except for the turbulent inflow field.
For more technical details, we refer to \citep[ suppl. mat.]{Engels2015}.

\subsection{Inflow turbulence\label{subsec:Inflow-turbulence}}

Flying animals encounter a considerable variety of aerial perturbations
while foraging, ranging from no perturbation in almost quiescent air
when the weather is calm, to fully turbulent, with intermittent gusts
and vortices generated by obstacles, such as flowers, trees or buildings.
The type of perturbation also depends on behavioral patterns in animals.
Bees, for example, forage on flowers and thus regularly perform landing
maneuvers which force them to fly in the flower's wake. Owing to this
huge variability in turbulent perturbations, we first reduce the parameter
space. Therefore we define a typical turbulent flow and choose homogeneous
isotropic turbulence (HIT) for the upstream perturbations because
it is the most widely used. It is also realized in experimental work,
e.g., generated by a grid in a wind tunnel \citep{Crall2016}. HIT
is characterized by its turbulent kinetic energy $E=3u_{\mathrm{RMS}}^{2}/2$,
or equivalently the turbulence intensity $I=u_{\mathrm{RMS}}/u_{\infty}$,
the Reynolds number $Re_{\lambda}=u{}_{\mathrm{RMS}}\lambda/\nu$,
based on the Taylor-micro scale $\lambda=\sqrt{15\nu u_{\mathrm{RMS}}/\varepsilon}$,
and the integral length scale 
\[
\Lambda=\frac{\pi}{2u_{\mathrm{RMS}}^{2}}\int_{k>0}^{\infty}k^{-1}E(k)\,\mathrm{d}k\,.
\]
Here, $\nu$ is the kinematic viscosity, $\varepsilon$ the dissipation
rate, $k$ the wavenumber and $E(k)$ is the energy spectrum integrated
over wavenumber shells. Note that for spatially periodic velocity
fields, the integral reduces to a sum, as only integer wavenumbers
$k\in\mathbb{N}$ exist. We pre-compute the HIT velocity fields in
a separate direct numerical simulation. In this computation, energy
is injected at a given wavenumber $k_{f}$ to compensate for the loss
due to viscous dissipation. Forced wavenumbers in the shell $k_{f}-0.5\leq\left|k\right|\leq k_{f}+2.5$
are multiplied with a factor $c(t)$ to keep the overall energy constant
in time. This approach is known as negative viscosity forcing \citep{Jimenez1993,Ishihara2003}.
In all HIT computations, we resolve the Kolmogorov scale $\eta=\left(\nu^{3}/\epsilon\right)^{1/4}$,
hence $k_{\mathrm{max}}\eta\gtrsim1$. We start the HIT simulations
with a random initial condition with prescribed spectrum \citep{Rogallo1981}.
After the statistically steady state has been reached, we save velocity
fields for later use as inflow perturbations. The saving interval
is at least 10 eddy turnover times to assure that the fields are uncorrelated
in time. By modifying $k_{f}$ at constant $E$ and $\nu$, we vary
the spectral distribution of energy.

\begin{table}[h]
\noindent \begin{centering}
\begin{tabular}{cccccccccc}
\toprule 
\multirow{2}{*}{Series} & \multirow{2}{*}{$I$} & \multirow{2}{*}{$\Lambda\,[R]$} & \multirow{2}{*}{$\lambda\,[R]$} & \multirow{2}{*}{$\eta\,[R]$} & \multirow{2}{*}{$T_{0}\,[T]$} & \multirow{2}{*}{$Re_{\lambda}$} & \multirow{2}{*}{$k_{\mathrm{max}}\eta$} & \multicolumn{2}{c}{$N_{R}$}\tabularnewline
\cmidrule{9-10} 
 &  &  &  &  &  &  &  & tethered & free\tabularnewline
\midrule
\midrule 
\multirow{4}{*}{A} & 0.16 & 0.77 & 0.25 & 0.013 & 3.67 & 90 & 1.72 & 4 & 3\tabularnewline
\cmidrule{2-10} 
 & 0.33 & 0.77 & 0.18 & 0.008 & 0.19 & 129 & 1.07 & 5 & 16\tabularnewline
\cmidrule{2-10} 
 & 0.60 & 0.76 & 0.13 & 0.005 & 0.98 & 177 & 0.99 & 9 & 9\tabularnewline
\cmidrule{2-10} 
 & 0.99 & 0.76 & 0.11 & 0.004 & 0.62 & 227 & 0.94 & 27 & 6\tabularnewline
\midrule 
\multirow{3}{*}{B} & 0.33 & 1.54 & 0.26 & 0.01 & 3.62 & 186 & 1.32 & $10^{(a)}$ & --\tabularnewline
\cmidrule{2-10} 
 & 0.33 & 0.77 & 0.18 & 0.008 & 1.91 & 129 & 1.07  & 5 & 16\tabularnewline
\cmidrule{2-10} 
 & 0.33 & 0.32 & 0.11 & 0.006 & 0.77 & 82 & 1.70 & 5 & 15\tabularnewline
\bottomrule
\end{tabular}
\par\end{centering}
$^{(a)}$ {\footnotesize{}We computed two runs with four insects and
an additional two runs with only one.}{\footnotesize\par}
\noindent \centering{}\caption{Properties of the homogeneous isotropic turbulence fields (time averaged
over several eddy-turnover times) used as inflow perturbations for
the insect. The rightmost column shows the number of realizations
$N_{R}$ used in tethered- and free flight simulations. Here, $I$
is the turbulence intensity, which is equivalent to the turbulent
kinetic energy, $\Lambda$ is the integral scale, $\lambda$ is the
Taylor microscale, $\eta$ the Kolmogorov scale and $T_{0}$ is the
eddy turnover time. All quantities are given in units of wing length
$R$ and wing beat duration $T$. \label{tab:Properties-of-inflow}}
\end{table}

\begin{figure}[h]
\noindent \begin{centering}
\includegraphics[width=0.5\textwidth]{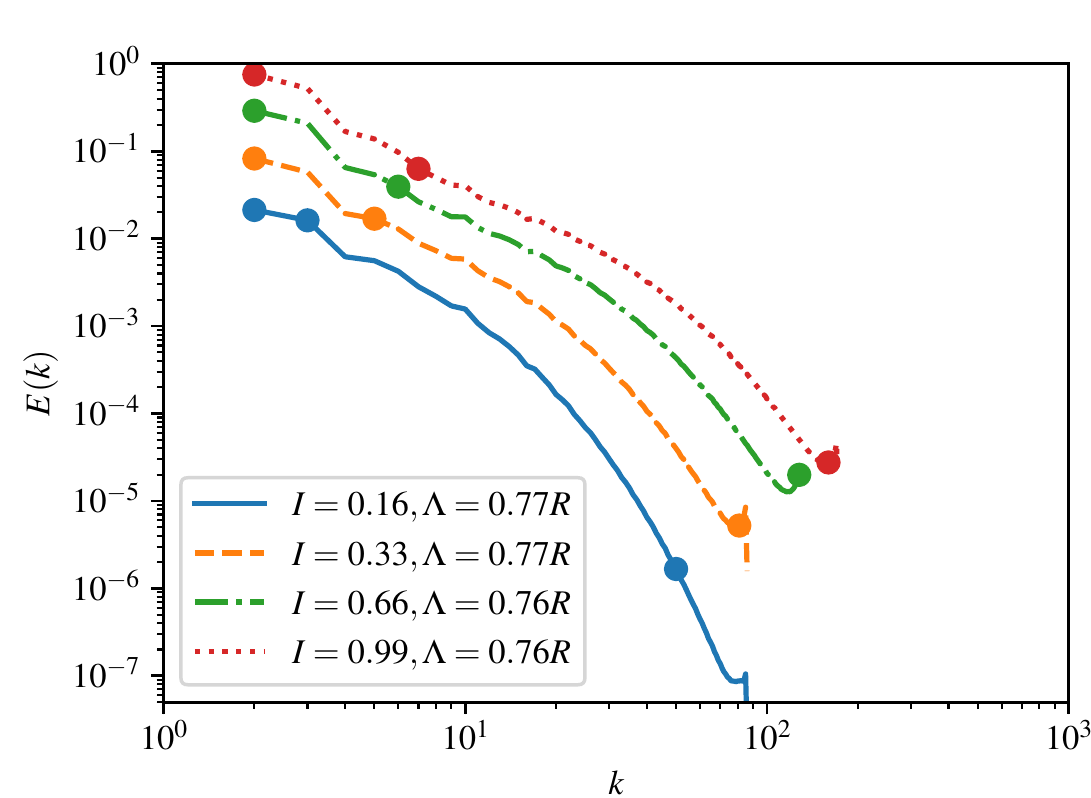}\includegraphics[width=0.5\textwidth]{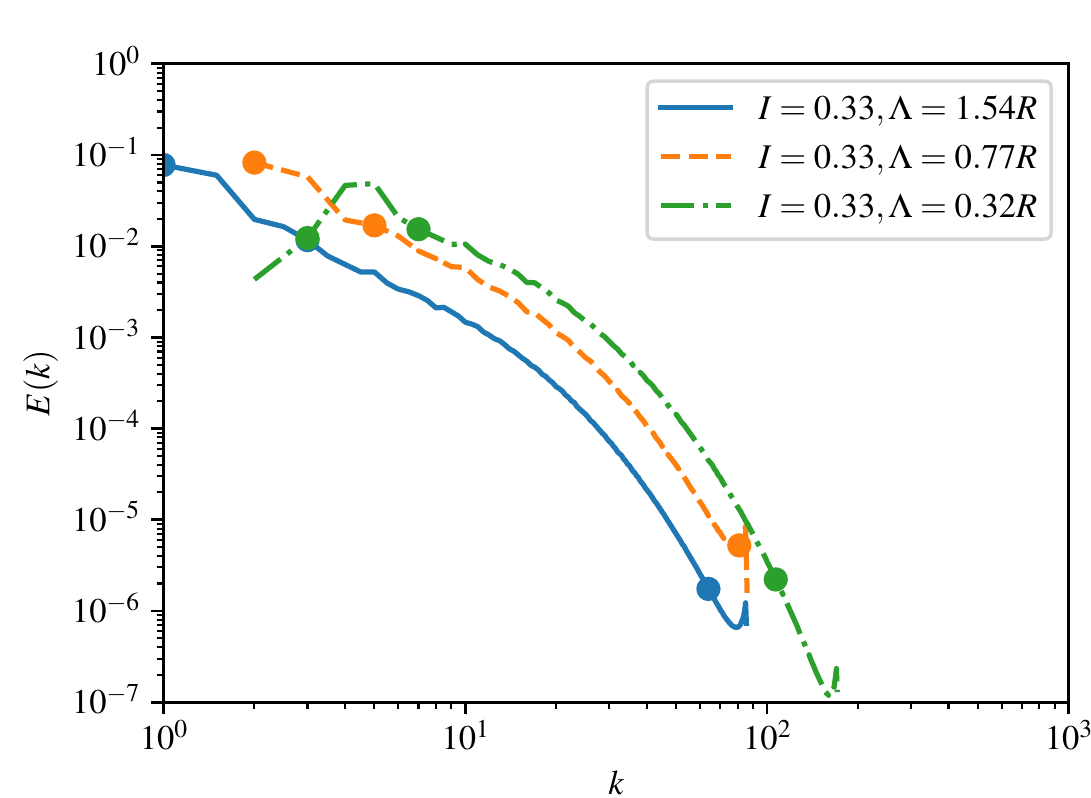}
\par\end{centering}
\caption{Spectra of turbulent kinetic energy for HIT fields averaged over several
eddy turnover times. Left: series A with constant $\Lambda$ and variable
$I$, right: series B for $I=0.33$ and $\Lambda$ variable. Markers
are the wavenumbers associated with $\Lambda$, $\lambda$ and $\eta$
(from left to right on each spectrum). \label{fig:Energy-spectra-HIT} }
\end{figure}

We generate two series of HIT simulations with turbulent kinetic energy
spectra shown in Fig. \ref{fig:Energy-spectra-HIT}. In series A we
vary the intensity $I$ from mildly ($I=0.16$) to extremely ($I=0.99$)
turbulent while keeping the integral length scale $\Lambda=0.77R$
fixed (Fig. \ref{fig:Energy-spectra-HIT} left). In series B where
we fix $I=0.33$ and vary $\Lambda=\{0.32R,\,0.77R,\,1.54R\}$ (Fig.
\ref{fig:Energy-spectra-HIT} right). Turbulence properties are summarized
in Table \ref{tab:Properties-of-inflow}. The first series allows
us to evaluate the impact of turbulence intensity, while the second
series allows us to investigate the influence of $\Lambda$ on the
insect. Note that the eddy turnover time $T_{0}=\Lambda/u_{\mathrm{RMS}}$
decreases, as expected, with increasing $I$ (series A) and, likewise,
with decreasing $\Lambda$ (series B). We vary the energy distribution
\emph{via} the forcing wavenumber $k_{f}$ in the HIT simulation.
Note that in the $\Lambda=0.77R$ case the forcing wavenumber was
$k_{f}=1$, thus we cannot reduce it any further in order to increase
$\Lambda$. Therefore, in order to increase $\Lambda$ to $1.54R$,
we double the lateral domain size to $L_{y}=L_{z}=8R$ in the insect
simulation, which then allows $k_{f}=1$ to result in a larger integral
scale. With the larger domain, we then compute four identical insects
in one simulation (Fig. \ref{fig:Setup-detailled-description}D),
to reduce the computational cost.

\begin{figure}[h]
\noindent \begin{centering}
\includegraphics[width=0.8\textwidth]{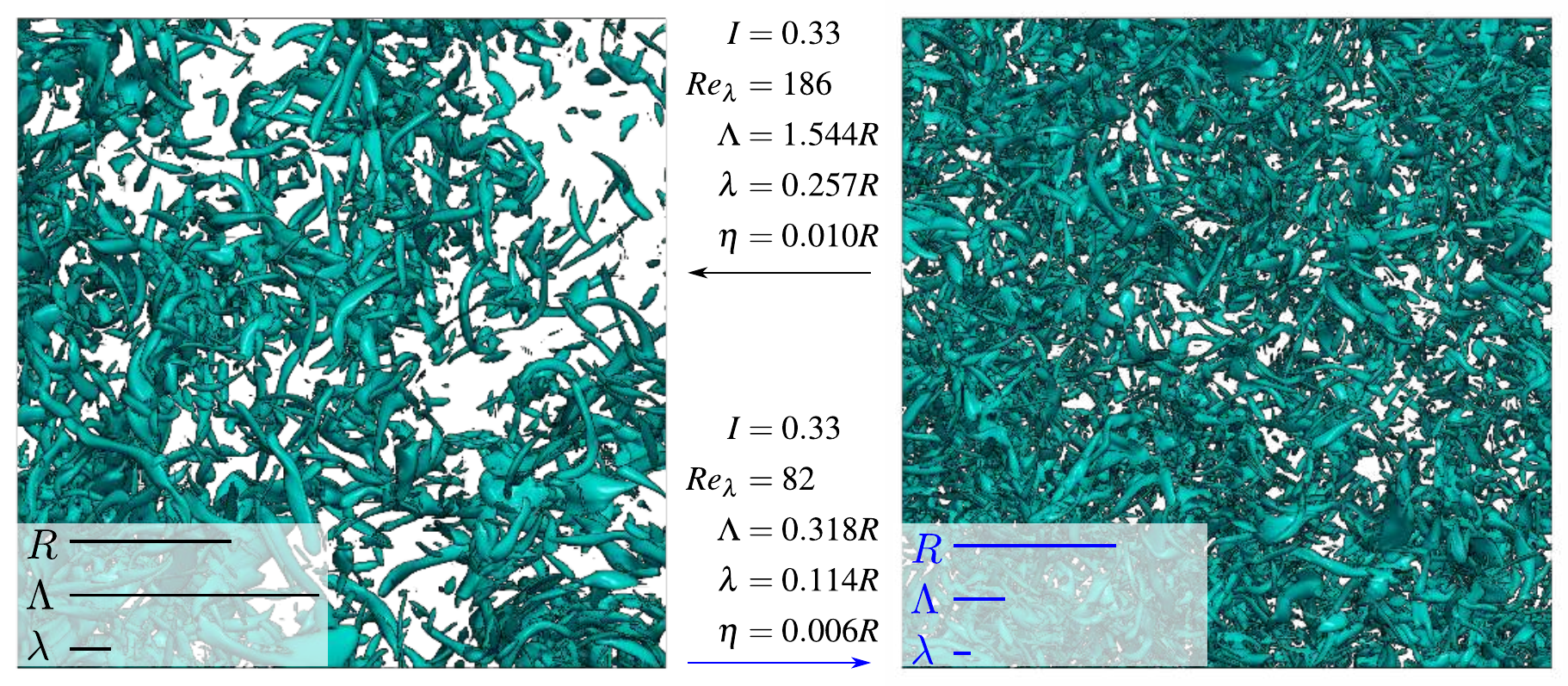}
\par\end{centering}
\caption{\label{fig:HIT-visu}Two HIT fields from the series B, both with identical
turbulent kinetic energy (and hence turbulence intensity $I$). Visualized
is the isosurface $\left|\underline{\omega}\right|=4\sigma(\underline{\omega})$
of vorticity magnitude, where $\sigma$ is the standard deviation.
Insets show visual comparison of wing length $R$, integral scale
$\Lambda$ and Taylor micro scale $\lambda$. }
\end{figure}

Fig. \ref{fig:HIT-visu} illustrates two individual HIT fields from
the B series by showing the isosurface of vorticity $\left|\underline{\omega}\right|=4\sigma(\underline{\omega})$,
where $\sigma$ is the corresponding standard deviation. The energy
$E$ of both fields is the same, but the integral scales are $\Lambda=1.54R$
and $0.32R$, respectively. Visibly, the $\Lambda=0.32R$ case features
smaller scale vortices which are more densely distributed in the periodic
box.

\subsection{Bumblebee model\label{subsec:Bumblebee-model}}

In our numerical simulations, we use a model bumblebee in forward
flight at $u_{\infty}=2.5\,\mathrm{m/s}$ as archetype for medium-sized
insects. The Reynolds number is $\mathrm{Re}=\overline{U}_{\mathrm{tip}}c_{m}/\nu_{\mathrm{air}}=2060$,
where $\overline{U}_{\mathrm{tip}}=2\Phi Rf=8.05\,\mathrm{m/s}$ is
the mean wingtip velocity, $c_{m}=4.012\,\mathrm{mm}$ the mean chord
length, $\nu_{\mathrm{air}}=1.568\cdot10^{-5}\,\mathrm{m^{2}/s}$
is the kinematic viscosity of air, $R=1.32\cdot10^{-2}\mathrm{\,m}$
is the wing length, $f=152\,\mathrm{Hz}$ ($T=1/f=6.6\,\mathrm{ms}$)
is the wingbeat frequency ($T$ is duration) and $\Phi=115^{\circ}$
is the wingbeat amplitude. The model is described in greater detail
elsewhere \citep[ suppl. mat.]{Engels2015}. The mass of the insect
is $m=175$ mg, the gravitational acceleration $g=9.81\,\mathrm{m/s^{2}}$
and the moments of inertia of the body are $J_{\mathrm{roll}}^{(b)}=1.14\cdot10^{-9}\,\mathrm{kg\,m^{2}}$,
$J_{\mathrm{yaw}}^{(b)}=4.33\cdot10^{-9}\,\mathrm{kg\,m^{2}}$ and
$J_{\mathrm{pitch}}^{(b)}=4.18\cdot10^{-9}\,\mathrm{kg\,m^{2}}$.
We use the superscript $\cdot^{(b)}$ when referring to the body reference
frame.

\begin{figure}[h]
\noindent \begin{centering}
\includegraphics[width=0.85\textwidth]{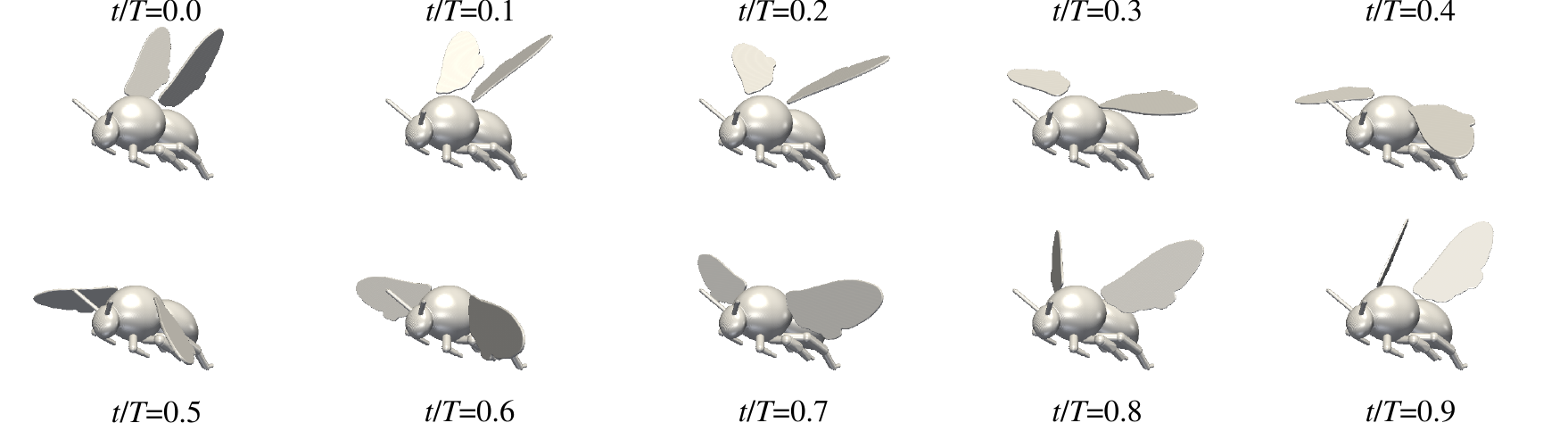}
\par\end{centering}
\caption{Visualization of the bumblebees prescribed flapping motion every $0.1T$
time steps. \label{fig:wingbeat-visualization}}
\end{figure}

We perform two types of simulations, one where the insect is anchored
to the virtual wind tunnel (tethered flight) and one where its motion
is computed from fluid forces and moments (free flight) as well as
gravity. The governing equation for the free flight case is Newton's
second law of motion for linear and angular motion. For the latter,
we use a quaternion Ansatz to avoid the Gimbal lock
problem. Gimbal lock occurs when two rotation axis become parallel
and the system looses one degree of freedom. The detailed set of
13 first-order ODEs can be found in \citep{Engels2015a}. In both
free and tethered flight, we prescribe an identical wing motion relative
to the body, as illustrated in Fig. \ref{fig:wingbeat-visualization}.
The wing motion is identical for all wing beats. The wings and body
are assumed to be rigid.

Our bumblebee model responds in the free flight case, unlike real
animals, entirely passively to perturbations. Therefore, we limit
the simulation time to the order of magnitude of the reaction time
delay $\tau_{\mathrm{react}}$ in those animals. After this delay,
the insect may employ active steering mechanisms and modify the wing
beat or body posture. Previous studies on freely flying honeybees
reported response delays of approximately $20\,\mathrm{ms}$ or $4.5$
stroke cycles, suggesting the use of ocellar pathways for body stability
reflexes in this species \citep{Vance2013}. By contrast, recent work
\citep{Beatus2015} suggests reaction times of about $5\,\mathrm{ms}$
in fruit flies. The precise delay in bumblebees is unknown but expected
to be of the same order of magnitude as in honeybees. Therefore, we
simulate $8$ stroke cycles ($52.6\,\mathrm{ms}$) in a simulation,
allowing thus to quantify the response for any $\tau_{\mathrm{react}}\leq8T$.
Notably, we do not exactly know under which conditions insects react
at all to perturbations, or simply accept the externally imposed change
in flight direction and orientation. An example for this is shown
in \citep{Ravi2016}, where bumblebees are found to ignore aerial
perturbations when approaching a cylinder.

\subsection{Governing equations and numerical method\label{subsec:Governing-equations-and}}

The present work relies on numerical simulations. We directly solve
the incompressible Navier--Stokes equations without any \emph{a priori}
turbulence models. All scales of fluid motion are fully resolved in
time and space. In this section, we describe briefly the numerical
method we use, for reasons of self-consistency. For further details
the reader is referred to \citep{Engels2015a}.

We employ a Fourier pseudospectral method for spatial discretization
and a 2nd order Adams-Bashforth scheme for time advancement. The spectral
discretization is fast and accurate \citep{Peyret2002} and is particularly
useful in our case as the Laplace operator becomes diagonal in Fourier
space. Hence, the solution of a Poisson problem is trivial in Fourier
space. To include the no-slip boundary conditions on the time-varying
geometry we use the volume penalization method \citep{Angot1999}.
This allows us to maintain the advantages of the Fourier discretization.
Hence, we solve the penalized Navier--Stokes equation
\begin{eqnarray}
\partial_{t}\underline{u}+\underline{\omega}\times\underline{u} & = & -\nabla\Pi+\nu\nabla^{2}\underline{u}-\underbrace{\frac{\chi}{C_{\eta}}\left(\underline{u}-\underline{u}_{s}\right)}_{\text{penalization}}-\underbrace{\frac{1}{C_{\mathrm{sp}}}\nabla\times\frac{\left(\chi_{\mathrm{sp}}\underline{\omega}\right)}{\nabla^{2}}}_{\text{sponge}}\label{eq:PNST_org_momentum}\\
\nabla\cdot\underline{u} & = & 0\label{eq:PNST_divergence_free}\\
\underline{u}\left(\underline{x},t=0\right) & = & \underline{u}_{0}\left(\underline{x}\right)\qquad\underline{x}\in\Omega,t>0,\label{eq:PNST_inicond}
\end{eqnarray}
where $\underline{u}$ is the fluid velocity, $\underline{\omega}=\nabla\times\underline{u}$
is the vorticity. We normalize the density $\varrho_{f}$ to unity.
The nonlinear term in eqn. (\ref{eq:PNST_org_momentum}) is written
in the rotational form. Hence we are left with the gradient of the
total pressure $\Pi=p+\frac{1}{2}\underline{u}\cdot\underline{u}$
instead of the static pressure $p$ \citep{Peyret2002}. This formulation
is chosen because of its favorable properties when discretized with
spectral methods, namely conservation of momentum and energy \citep[pp. 210]{Peyret2002}.
At the exterior of the computational domain, we assume periodic boundary
conditions. The domain is sufficiently large to minimize the effect
of periodicity.

The mask function $\chi$ is defined as
\begin{equation}
\chi\left(\underline{x},t\right)=\left\{ \begin{array}{cc}
0 & \text{if }\underline{x}\in\Omega_{f}\\
1 & \text{if }\underline{x}\in\Omega_{s}
\end{array}\right.,\label{eq:mask_function_discontinuous}
\end{equation}
where $\Omega_{f}$ is the fluid and $\Omega_{s}$ the solid domain.
Note that in the fluid domain $\Omega_{f}$, the original equations
hold  as the penalization term $\frac{\chi}{C_{\eta}}\left(\underline{u}-\underline{u}_{s}\right)$
vanishes. The convergence proof in \citep{Carbou2003,Angot1999} shows
that the solution of the penalized Navier--Stokes equations (\ref{eq:PNST_org_momentum}-\ref{eq:PNST_inicond})
tends for $C_{\eta}\rightarrow0$ indeed towards the exact solution
of Navier--Stokes imposing no-slip boundary conditions. Here, we
use $C_{\eta}=2.5\cdot10^{-4}$. We also add a second penalization
term for the vorticity $\underline{\omega}$, which we call sponge
term. The sponge gradually damps the vorticity in regions where $\chi_{\mathrm{sp}}=1$.
The sponge constant is set to $C_{\mathrm{sp}}=10^{-1}$.

In the case of free flight, we compute the position and orientation
of the insect from the aerodynamic forces and moments using a quaternion-based
formulation. We integrate the resulting ODE system time using the
same Adams-Bashforth scheme as for the fluid. More details about the
numerical method and its implementation in the open-source code \texttt{FluSI}
\footnote{Available free of charge and without registration on https://github.com/pseudospectators/FLUSI}
can be found in \citep{Engels2015a}, along with detailed validation
cases. In addition, appendix A shows the convergence
of the forces for decreasing wing thickness of a flapping wing.

\section{Results and discussion\label{sec:Results-and-discussion}}

In the following subsection, we present and discuss the results of
two types of simulations, tethered and free flight. We use both cases
to investigate the influence of turbulence on the insect when varying
either the intensity or the length scales of the turbulent inflow
perturbations. We start with the tethered cases, which serves as reference
for the free flight cases. In numerical simulations, the tethered
case is the idealized limit of perfect control. In experimental work,
where the animals are fixed using a material tether, usually a thin
wire glued to the back, the insects lack sensor feedback present in
free flight. The wing kinematics might then be very different from
what an insect uses in free flight \citep{Srygley2002}. However,
note that our tethered simulations are based on wing kinematics measured
in free flight \citep{Dudley1990,Dudley1990a}. They are thus equivalent
to a tethered insect that flaps as if it was in free flight.

\begin{figure}[h]
\noindent \begin{centering}
\includegraphics[width=0.23\textwidth]{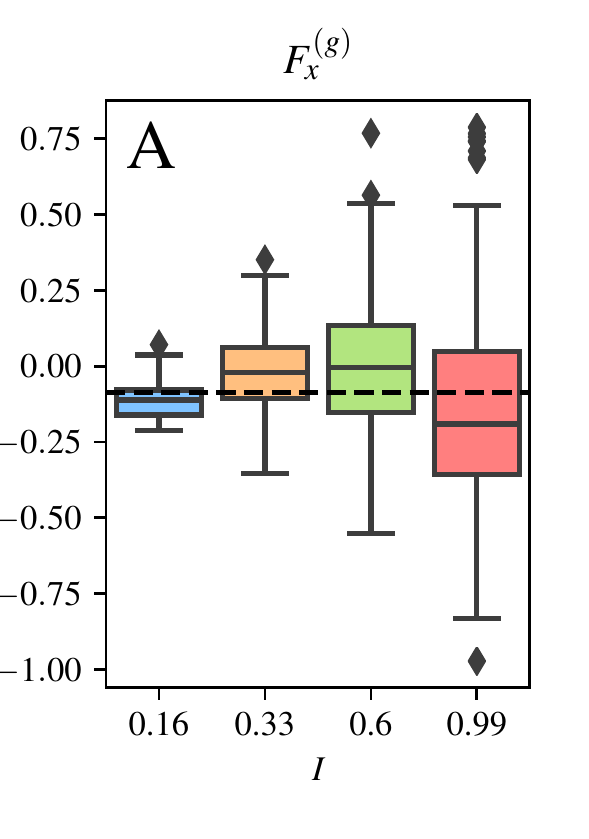}\includegraphics[width=0.23\textwidth]{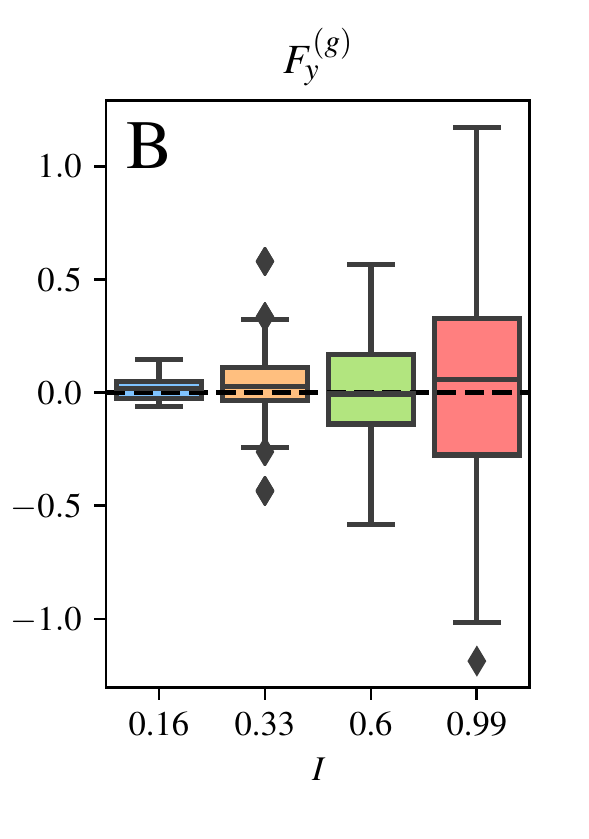}\includegraphics[width=0.23\textwidth]{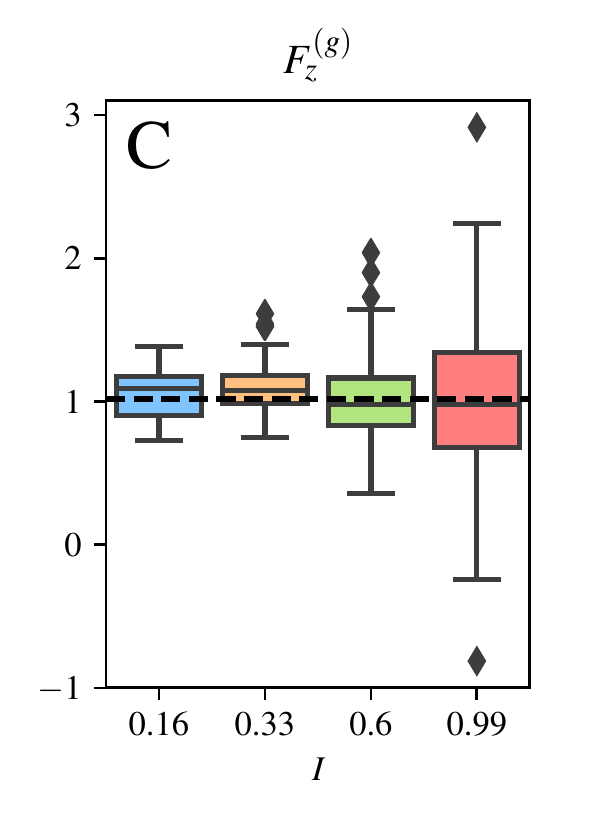}
\par\end{centering}
\noindent \begin{centering}
\includegraphics[width=0.23\textwidth]{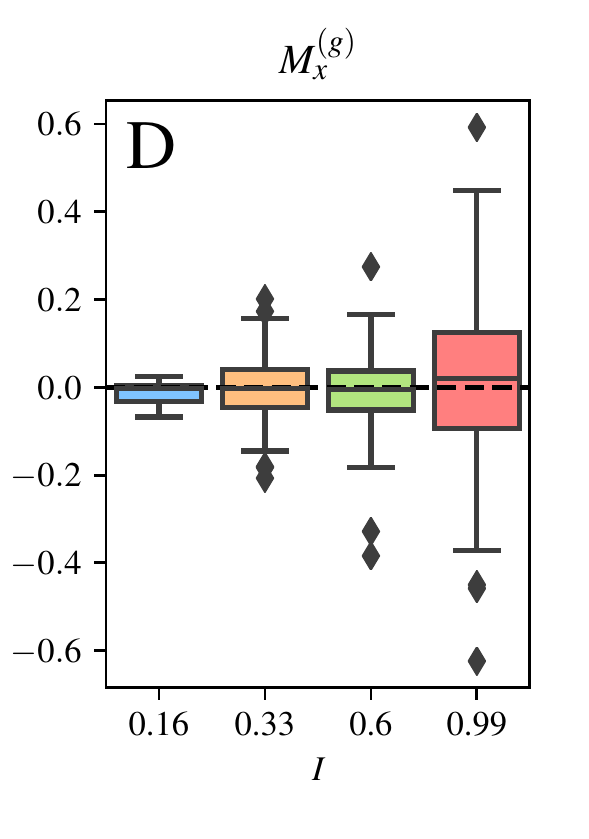}\includegraphics[width=0.23\textwidth]{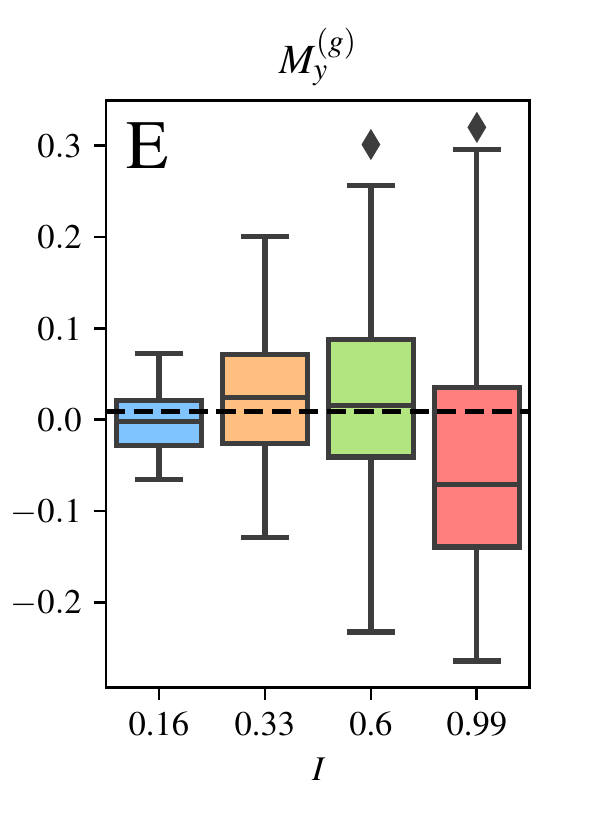}\includegraphics[width=0.23\textwidth]{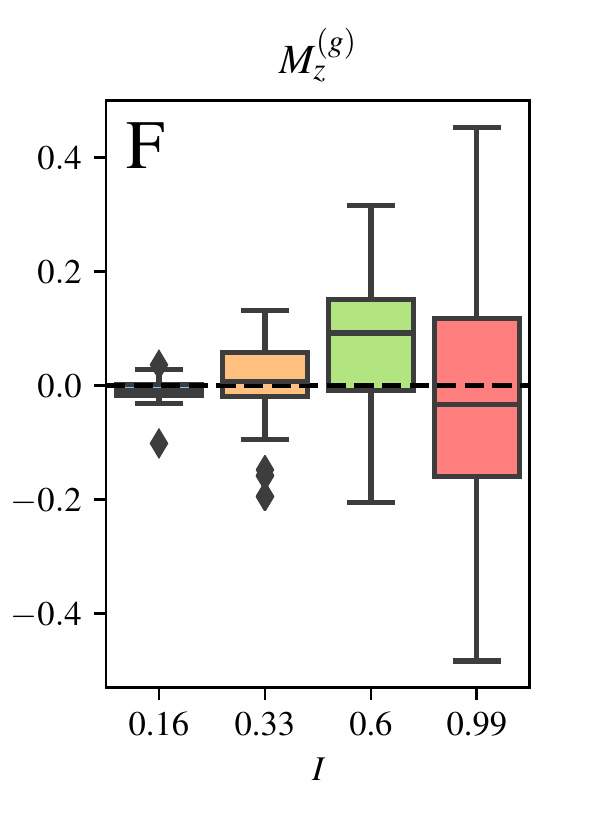}\includegraphics[width=0.23\textwidth]{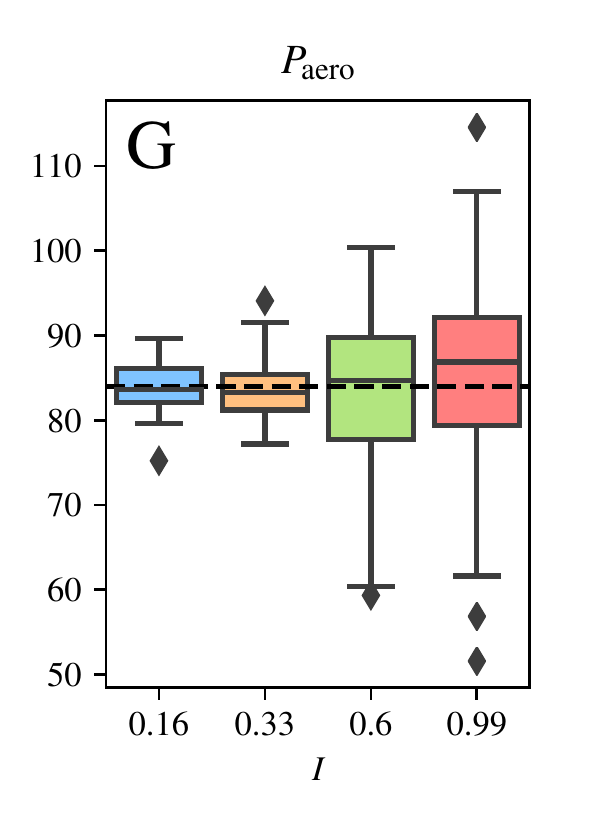}
\par\end{centering}
\caption{\label{fig:Tethered-turbulent-fluctuations-TU16-99}Tethered flight
in turbulence. The integral scale is $\Lambda=0.77R$ and $I$ varies
between $0$ and $0.99$. Cycle averaged values are represented by
box plots. Each of the $N_{R}$ simulations yields $4$ data points,
one for each cycle. In the colored boxes, the line is the median (or
2-quantile) of the data, and the limits of the box are the upper and
lower quartiles (or 4-quantile). The additional vertical line are
min/max values excluding outliers, which are shown as individual points
with a $\diamondsuit$ marker. Aerodynamic quantities are forces (A-C,
normalized by $mg$), moments (D-F, normalized by $mgR$) and aerodynamic
power (G, in W/kg body mass). The dashed line corresponds to the laminar
case ($I=0$) where the insect is aligned with the mean flow.}
\end{figure}

\subsection{Tethered flight}

\subsubsection{Influence of turbulence intensity at fixed length scales}

We first study the influence of $I$ at constant integral scale $\Lambda$.
The insect is tethered and we used the series A of turbulence fields,
as presented in \citep{Engels2015}. Their properties are summarized
in Table \ref{tab:Properties-of-inflow}. We fix the integral scale
and vary the turbulence intensity $I$, which also results in an
increasing Reynolds number $R_{\lambda}$ and reduced eddy turnover
time. Fig. \ref{fig:Tethered-turbulent-fluctuations-TU16-99} illustrates
the obtained results for forces (A-C), moments (D-F) and aerodynamic
power (G). We choose the box plot representation, first introduced
in \citep{Tukey1977}, to visualize the data. Each of the $N_{R}$
realizations yields $N_{w}$ independent cycle-averaged forces and
moments (Table \ref{tab:Properties-of-inflow}). The median value
of the data are remarkably close to the value in the laminar case
(dashed line) for all quantities, even in the strongest inflow turbulence.
This indicates that turbulence does not systematically alter the vortex
dynamics generated by the flapping wings of the insect. This vortex
system features the leading edge vortex that results from the typically
high angle of attack (here roughly $50^{\circ}$) \citep{Ellington1996,Ellington1999,Kolomenskiy2014a}.
The leading edge vortex remains attached to the wing in unperturbed
conditions, and in the simulations with turbulent inflow it is not
systematically detaching neither. Owing to the decreased pressure
in its core, this vortex provides a boost for the aerodynamic forces,
especially the lift force. Thus, its detachment or destruction would
result in a significant change in forces, moments and power. Compared
to an airfoil, where upstream turbulence can trigger transitions in
the boundary layer or impact flow separation, this behavior is thus
different. However, fluctuations occur, as represented in Fig. \ref{fig:Tethered-turbulent-fluctuations-TU16-99}
by the colored boxes and the min/max values. With increasing turbulence
intensity, those fluctuations become larger. We conclude that flapping
flight in turbulence faces insects more with a problem for control,
rather than deteriorated force production \citep{Engels2015}.

\subsubsection{Influence of turbulent length scales at constant intensity}

With the results of \citep{Engels2015} we now further explore the
influence of turbulent length scales on tethered flight and use the
series B from Table \ref{tab:Properties-of-inflow}, where we fixed
$I=0.33$. This particular intermediate value of $I$ does not require
a large number of flow realizations for any tested value of $\Lambda$,
which allows keeping the computational cost within acceptable limits.
Furthermore, field experiments \citep{Crall2016} show a large flight
activity of bumblebees for this value of $I$.

\begin{figure}[h]
\noindent \begin{centering}
\includegraphics[width=0.23\textwidth]{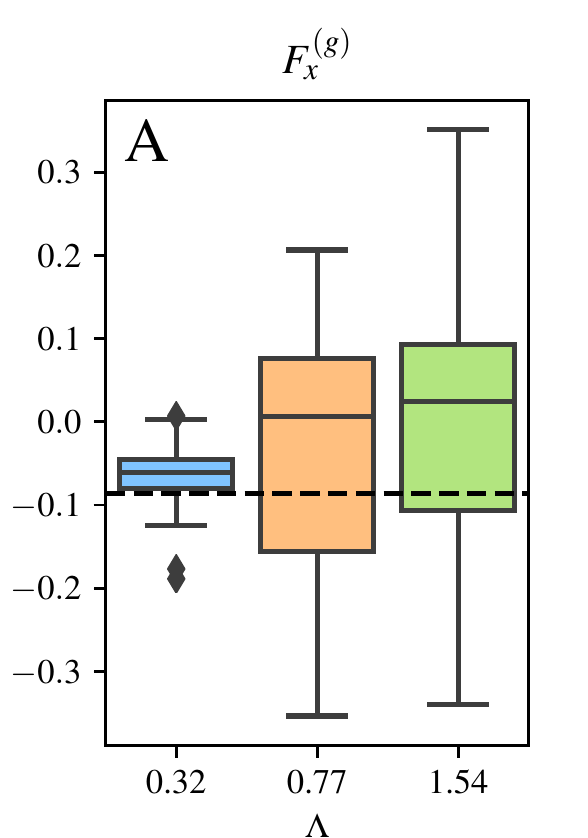}\includegraphics[width=0.23\textwidth]{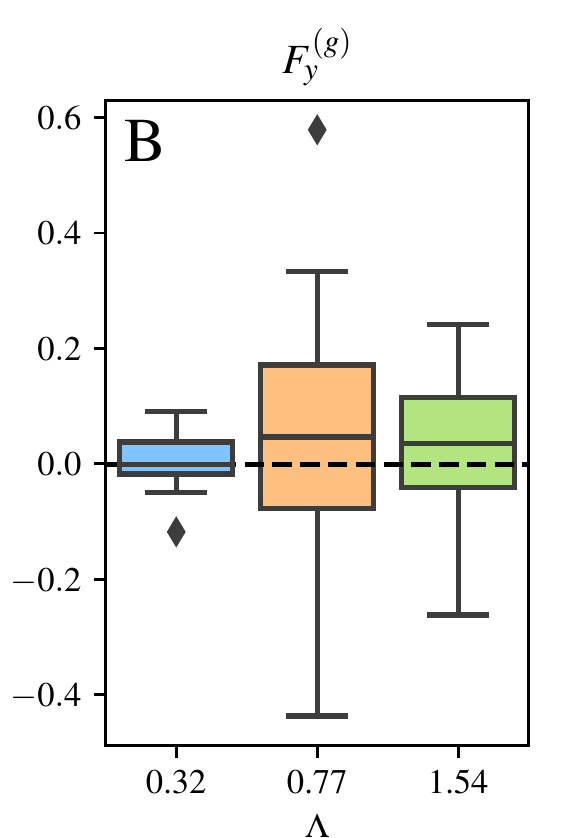}\includegraphics[width=0.23\textwidth]{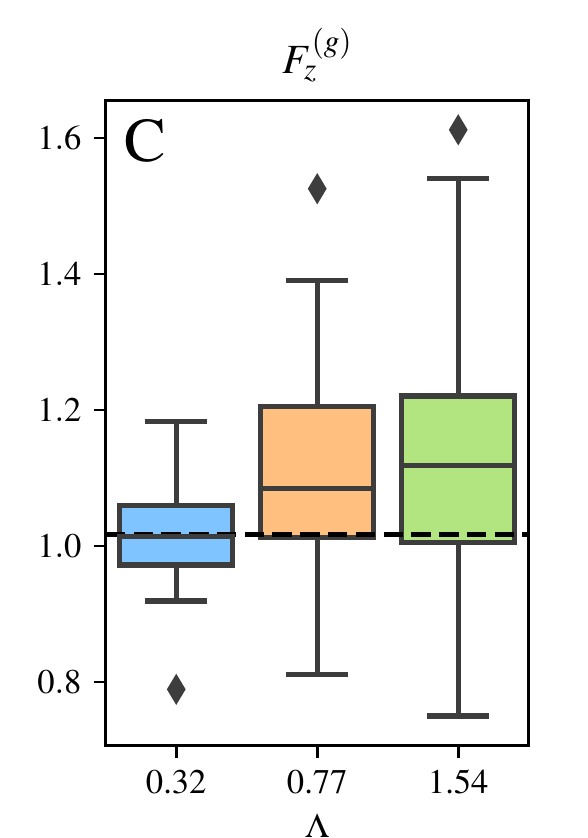}
\par\end{centering}
\noindent \centering{}\includegraphics[width=0.23\textwidth]{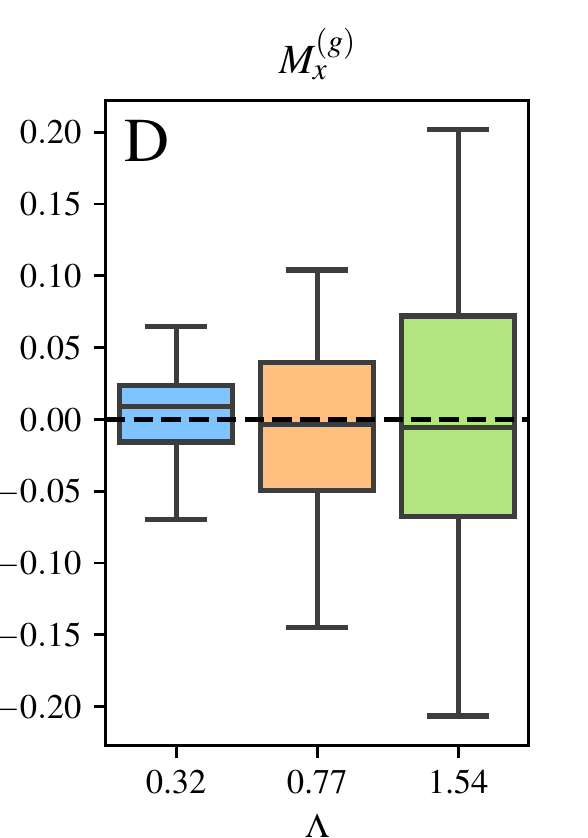}\includegraphics[width=0.23\textwidth]{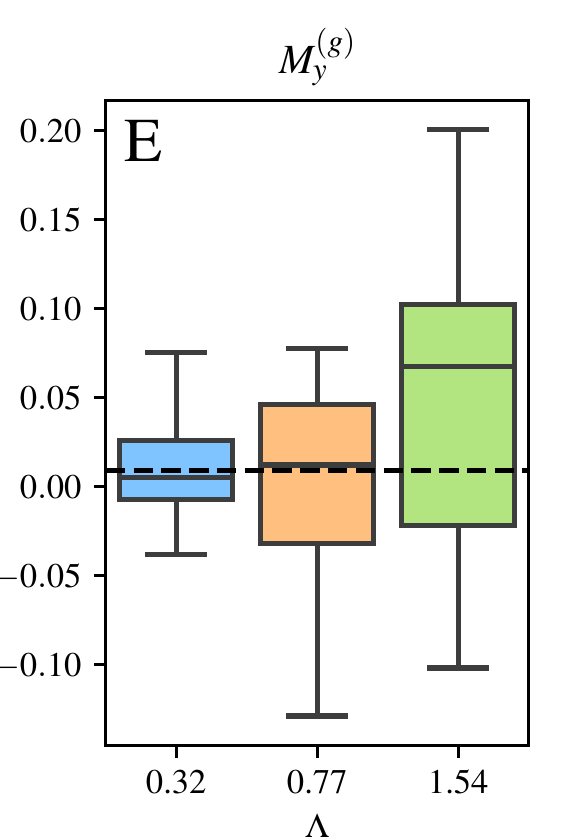}\includegraphics[width=0.23\textwidth]{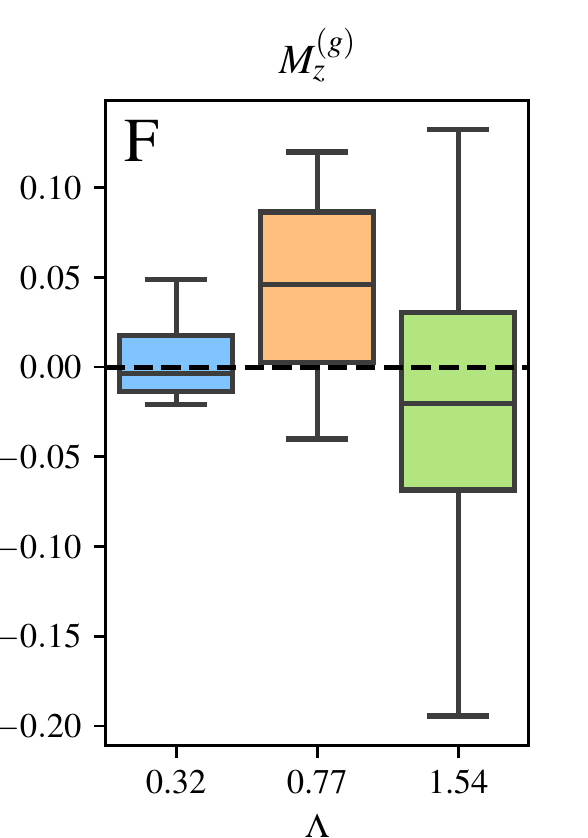}\includegraphics[width=0.23\textwidth]{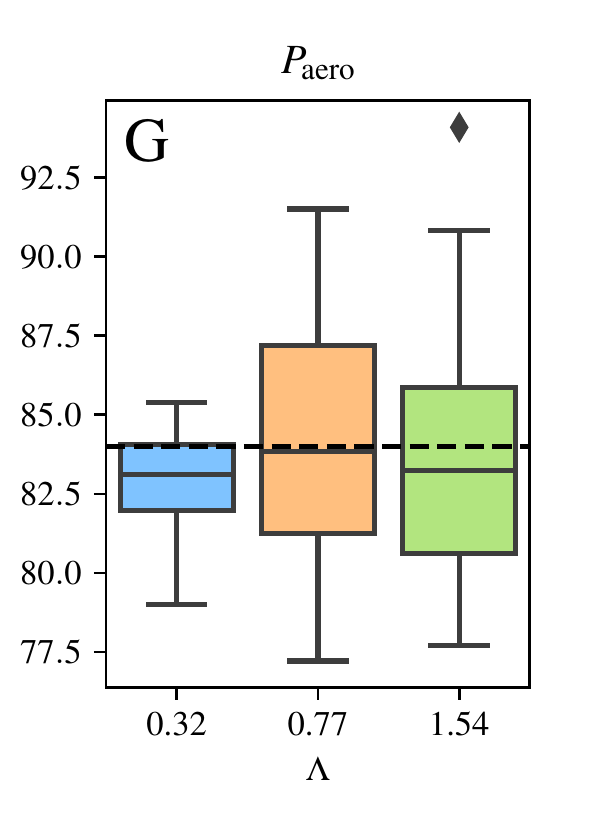}\caption{\label{fig:Tethered-turbulent-fluctuations-TU33-lambda}Tethered flight
in turbulence. The turbulence intensity is $I=0.33$ and the integral
scale $\Lambda$ varies between $0.32R$ and $1.54R$. Cycle averaged
values are represented by box plots. Each of the $N_{R}$ simulations
yields $4$ data points, one for each cycle. In the colored boxes,
the line is the median of the data, and the limits of the box are
the upper and lower quartiles. The additional vertical line are min/max
values excluding outliers, which are shown as individual points with
a $\diamondsuit$ marker. Aerodynamic quantities are forces (A-C,
normalized by $mg$), moments (D-F, normalized by $mgR$) and aerodynamic
power (G, in W/kg body mass). The dashed line corresponds to the laminar
case ($I=0$) where the insect is aligned with the mean flow.}
\end{figure}

Fig. \ref{fig:Tethered-turbulent-fluctuations-TU33-lambda} illustrates
the cycle-averaged forces, moments and power as a function of $\Lambda$.
The median values are close to the values in laminar inflow (dashed
line), which is consistent with the findings in \citep{Engels2015}
and Fig. \ref{fig:Tethered-turbulent-fluctuations-TU16-99}. For any
quantity, fluctuations are significantly reduced at $\Lambda=0.32R$
(blue), compared to the other two cases. The lateral (Fig. \ref{fig:Tethered-turbulent-fluctuations-TU33-lambda}B)
and lift (Fig. \ref{fig:Tethered-turbulent-fluctuations-TU33-lambda}C)
force exhibit the largest fluctuations for $\Lambda=0.77R$, while
the fluctuations in thrust (Fig. \ref{fig:Tethered-turbulent-fluctuations-TU33-lambda}A)
are of the same magnitude in both cases. For the aerodynamic torques
(Fig. \ref{fig:Tethered-turbulent-fluctuations-TU33-lambda}D-F) largest
fluctuations appear for $\Lambda=1.54R$ with standard deviation $\sigma=0.104$,
0.095, 0.076 for the roll ($M_{x}$), pitch ($M_{y}$) and yaw ($M_{z}$)
moments, respectively. The yaw moment is slightly less sensitive to
perturbations but remains of the same order of magnitude. The aerodynamic
power $P_{\mathrm{aero}}$ (Fig. \ref{fig:Tethered-turbulent-fluctuations-TU33-lambda}G)
displays the same behavior as the forces, with $\Lambda=0.77R$ resulting
in the largest fluctuations. However, in that case, $\sigma(P_{\mathrm{aero}})/\overline{P}_{\mathrm{aero}}=0.05$,
while for the vertical force $\sigma(F_{z})/\overline{F}_{z}=0.2$.
The power thus fluctuates little.

\begin{figure}[h]
\noindent \begin{centering}
\includegraphics[width=0.8\textwidth]{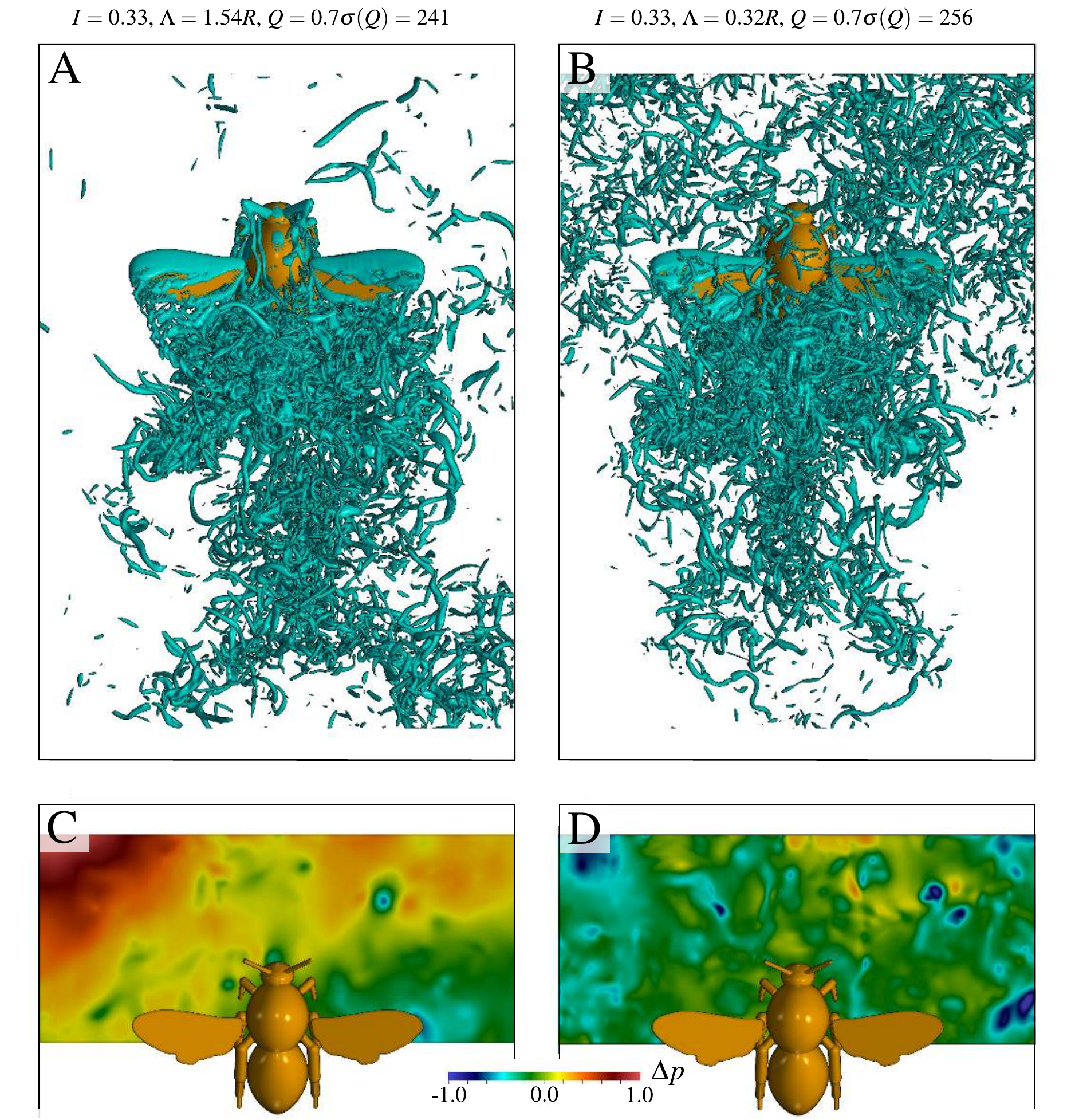}
\par\end{centering}
\caption{\label{fig:Tethered-flow-visualization}Tethered flight in turbulence.
Left column: $I=.33$ and $\Lambda=1.54R$, right column: $I=.33$
and $\Lambda=0.32R$. (A-B) An isosurface of the $Q$-criterion is
shown to visualize vortical structures. In the case of $\Lambda=1.54R$
(A), fewer vortices are identified upstream of the insect. (C-D) show
the corresponding pressure field, where the pressure from the laminar
inflow has been subtracted, $\Delta p=p_{\mathrm{turb}}-p_{\mathrm{lam}}$.
Compared to the $\Lambda=0.32R$ case (D), variations in pressure
are of the same order of magnitude but on a larger spatial scale in
the $\Lambda=1.54R$ case (C).}
\end{figure}

These results suggest a reduced sensitivity to turbulence at smaller
scales, expressed in a reduction of more than a factor of two in the
magnitude of fluctuations at the same turbulence intensity. This is
in agreement with the conjecture stated in \citep{Ravi2013} that
perturbations which are small compared to the animal average out over
the body and thus induce less perturbations. To further explore the
effect of $\Lambda$, we illustrate in Fig. \ref{fig:Tethered-flow-visualization}A-B
the flow for the coarsest and finest turbulent case. Vortical structures
are visualized by the Q-criterion \citep{Hunt1988}. For both inflow
conditions, we plot the same relative isosurface using the standard
deviation $\sigma$, $Q=0.7\sigma(Q)$, to identify vortices. In the
coarser turbulence, less vortex tubes can be identified in the region
between the inlet and the insect than in the smaller scale case, and
the tubes are of similar diameter. This may lead to the visual intuition
that the smaller scale turbulence has a larger impact on the insect.
However, the pressure field, illustrated in Fig. \ref{fig:Tethered-flow-visualization}C-D
as the difference in pressure between the turbulent and laminar realization,
$\Delta p=p_{\mathrm{turb}}-p_{\mathrm{lam}}$, confirms that pressure
fluctuations are of similar magnitude in both cases, while the spatial
scale differs significantly. The coarser scale turbulence is associated
with much larger scales of the pressure variations, which therefore
have less chance of canceling out over the region of the insect.

\subsection{Free flight}

We now consider our model in free flight with all six degrees of freedom
coupled to the flow solver, neglecting active control. This configuration
is more realistic for real insects, since they cannot react instantaneously
to changes in the flow condition. Reaction rather takes place after
a time delay $\tau_{\mathrm{react}}$, during which sensor information
are converted to changes in wing beat for active countermeasures (see
section \ref{subsec:Bumblebee-model}). Therefore, the insect behaves
passively during this interval, similar to what our model does. The
orientation and linear/angular velocities after $\tau_{\mathrm{react}}$
can thus yield insight into the effort required for corrective maneuvers.

\subsubsection{Influence of turbulence intensity at constant length scales}

After revisiting the problem of a tethered bumblebee in turbulence
and studying the same model in free flight and laminar inflow, we
now turn to free flight in turbulence. We first keep $\Lambda=0.77R$
fixed for these simulations and alter the energy content of the imposed
velocity fluctuations (series A in Table \ref{tab:Properties-of-inflow}).
In free flight, force and moment fluctuations are transduced to linear
and angular velocities, which in turn alter the forces and moments.
It can be seen as the limiting case of no flight control, while tethered
flight can somehow be seen as limit of perfect control using external
force, in the sense that attitude is perfectly stabilized while neglecting
the necessary changes in wing beat.

\begin{figure}[h]
\noindent \begin{centering}
\includegraphics[width=0.8\textwidth]{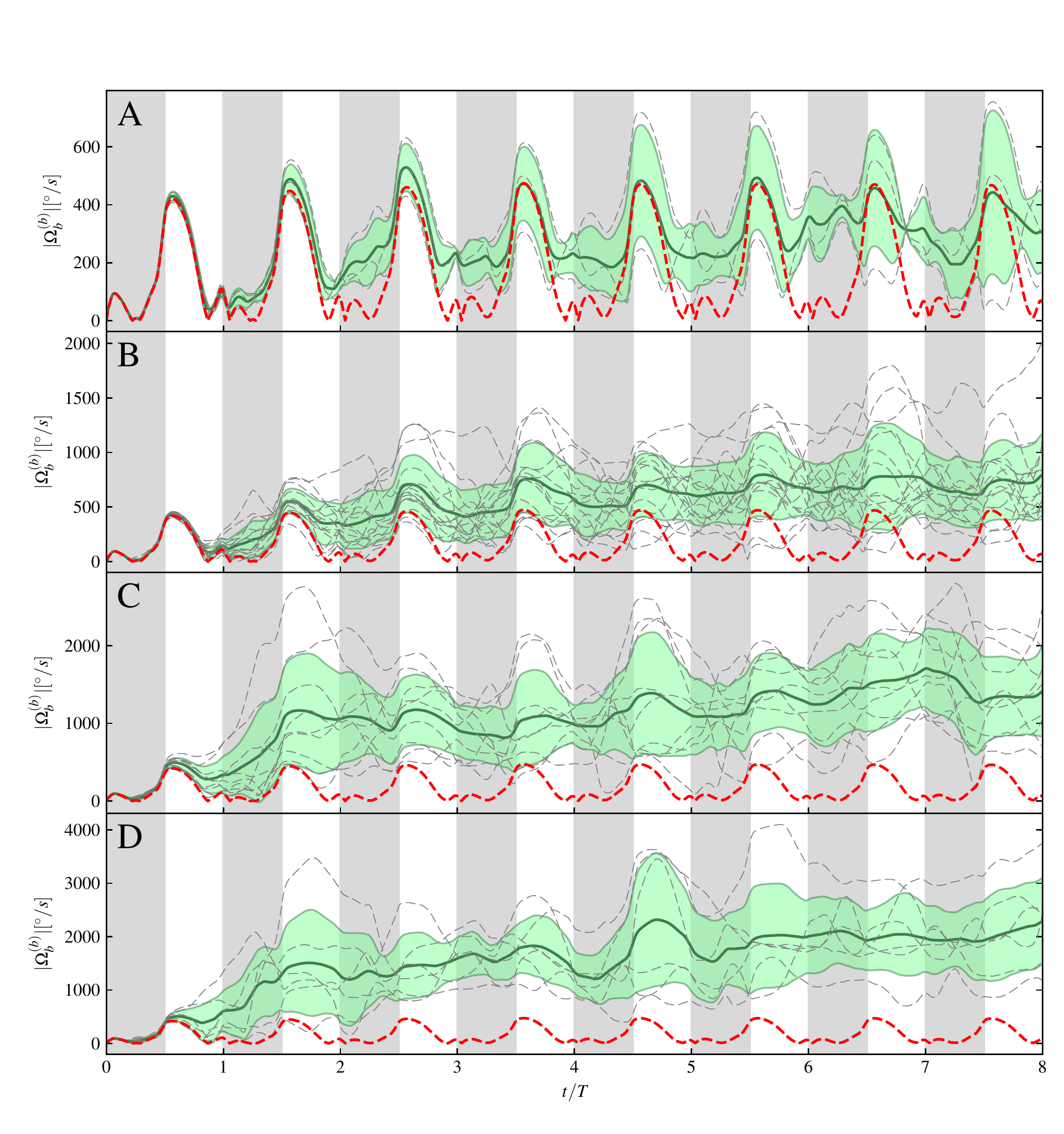}
\par\end{centering}
\caption{\label{fig:FF_tu016-99}Free flight in turbulent inflow, $I=0.16$
(A), $0.33$ (B), $0.60$ (C) and $0.99$ (D), for fixed $\Lambda=0.77$.
Time evolution for the magnitude of body angular velocity. Individual
realizations are shown as thin, gray, dashed lines. Ensemble averaged
time evolution is represented by the thick green line, light green
shaded background illustrates the standard deviation. The red dashed
line corresponds to the laminar case.}
\end{figure}

Fig. \ref{fig:FF_tu016-99}A-D shows the magnitude of the body's angular
velocity, $\underline{\Omega}_{b}^{(b)}$, as a function of time for
the four different turbulence intensities $I=0.16$, 0.33, 0.60 and
0.99. At the lowest turbulence intensity (Fig. \ref{fig:FF_tu016-99}A),
fluctuations remain small and the overall time evolution resembles
the laminar case, in which only the pitch component of $\underline{\Omega}_{b}^{(b)}$
is nonzero owing to lateral symmetry, although the difference grows
in time. The first stroke is virtually unaffected as perturbations
have not yet been advected to the insect. From the next larger value
of $I$ on (Fig. \ref{fig:FF_tu016-99}B), the resemblance to the
laminar case disappears. The terminal value of the ensemble averaged
angular velocity increases from $829^{\circ}/\mathrm{s}$ at $I=0.33$
(Fig. \ref{fig:FF_tu016-99}B) to $2300^{\circ}/\mathrm{s}$ at $I=0.99$
(Fig. \ref{fig:FF_tu016-99}D). In the laminar case, peak values of
$470^{\circ}/\mathrm{s}$ are found. It can be seen that after an
initial growth phase, which takes place roughly in the first two strokes,
the average angular velocity remains roughly constant, thus it is
limited by aerodynamic damping.

\begin{figure}[h]
\noindent \begin{centering}
\includegraphics[width=0.8\textwidth]{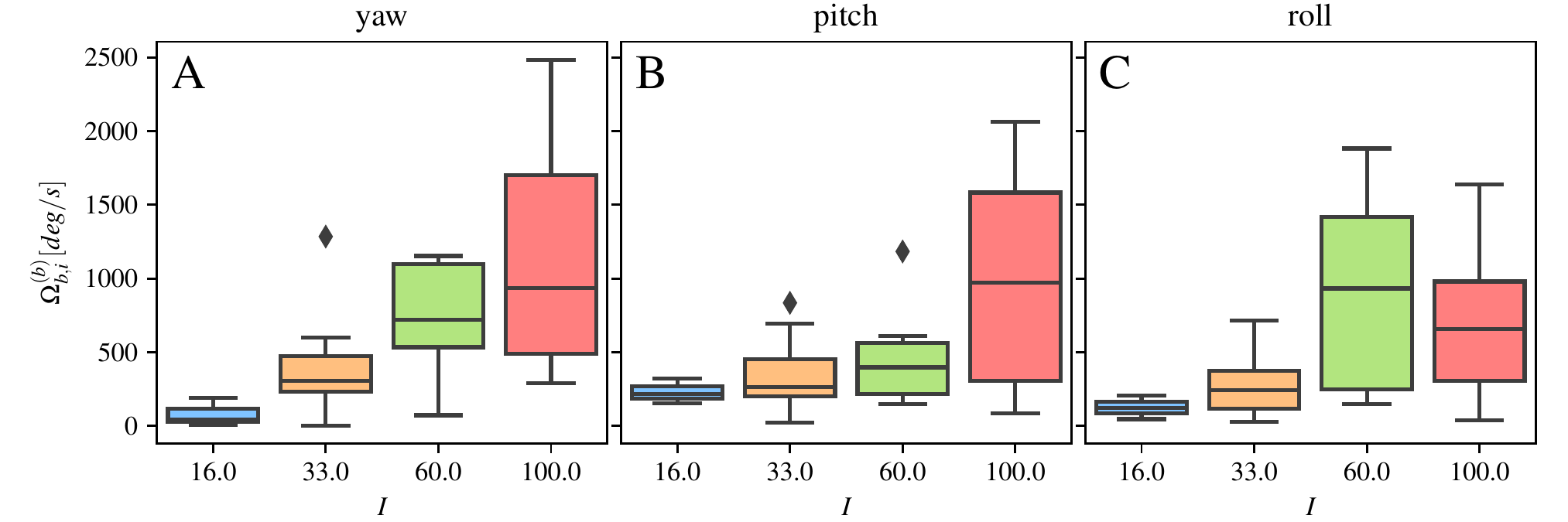}
\par\end{centering}
\caption{\label{fig:Free-flight-in-turbulent-TU-errorbars}Free flight in turbulent
inflow. Shown are the yaw (A), pitch (B) and roll (C) components of
the body angular velocity, averaged over the last computed cycle,
as a function of the turbulence intensity. Data are represented by
a box plot. In the colored boxes, the line is the median of the data,
and the limits of the box is the upper and lower quartiles. The additional
vertical line correspond to min/max values excluding outliers ($\diamondsuit$).}
\end{figure}

Fig. \ref{fig:Free-flight-in-turbulent-TU-errorbars} shows the components
of the angular velocity vector, averaged over the last cycle $7\leq t/T\leq8$,
as a function of the turbulence intensity. The magnitude of the mean
value as well as fluctuations increase with increasing $I$, but
no relevant difference among the three directions can be observed.
We thus do not observe a significantly increased roll angular velocity
(Fig. \ref{fig:Free-flight-in-turbulent-TU-errorbars}C), despite
the lower moment of inertia around this axis.

We find the largest magnitude of linear velocity $\left|\underline{u}_{b}^{(b)}\right|/u_{\infty}=0.06\pm0.04$
for the highest turbulence intensity ($I=0.99$, $\Lambda=0.77R$).
It can be concluded that, even for the largest turbulence intensity,
the translational response of the bumblebee is small compared to the
flight speed. Therefore, the changes in position $\underline{x}_{\mathrm{cntr}}$
are small within the time span of the computations, \emph{i.e.} $8T$.
The impact of turbulence on the angular degrees of freedom is thus
much higher than on the linear ones.

\subsubsection{Influence of turbulent length scales at constant intensity}

\begin{figure}[h]
\noindent \begin{centering}
\includegraphics[width=0.85\textwidth]{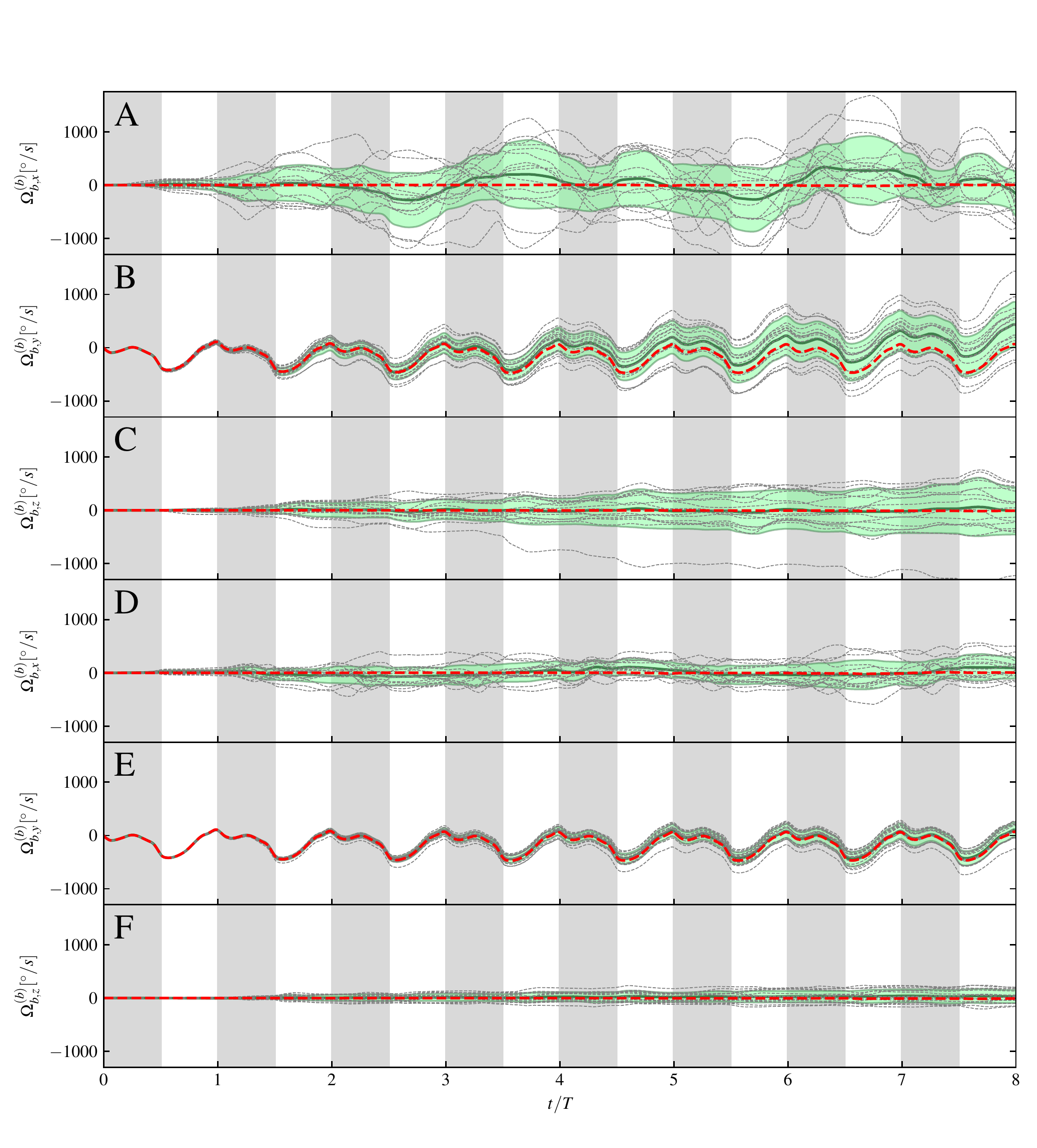}
\par\end{centering}
\caption{\label{fig:FF_tu033_lambda077_angles}Free flight in turbulent inflow,
$I=0.33$, A--C: $\Lambda=0.77R$, D--F: $\Lambda=0.32R$. Time
evolution of the three components of the angular velocity vector of
the body, in the body system $\underline{\Omega}_{b}^{(b)}$. Angular
velocities are given in $^{\circ}/s$ for easier comparison with results
in the literature. Individual realizations are shown as thin gray
lines. Ensemble averaged time evolution is represented by the thick
green line, light green shaded background illustrates the standard
deviation. The thick red dashed line corresponds to the laminar case.}
\end{figure}

As for the tethered case, we fix $I=0.33$ and vary the integral
scale $\Lambda$ of the turbulent inflow perturbations. Fig. \ref{fig:FF_tu033_lambda077_angles}
shows the angular velocity components for the case $I=0.33$, $\Lambda=0.77R$
(A--C) and $\Lambda=0.32R$ (D--F). Each realization is shown as
a thin gray line, the reference computation in laminar flow is shown
as red dashed line. All realizations result in a different attitude
of the insect, though the turbulence fields have identical statistical
properties. The ensemble-averaged time evolution (solid green lines)
is however remarkably close to what is seen in laminar inflow. Standard
deviations among the realizations (green shaded area) increase with
time, as turbulent inflow perturbations are imposed continuously. 

Ensemble averaged angular orientation, expressed in term of the body
angles, does not change significantly for $\psi_{\mathrm{roll}}$
and $\psi_{\mathrm{yaw}}$, while $\beta_{\mathrm{pitch}}$ has changed
by $8.5^{\circ}$. Fluctuations are largest for yaw ($15.4^{\circ}$),
followed by pitch ($11.0^{\circ}$) and roll ($10.3^{\circ}$). The
values are however close to each other, such that the difference is
not significant. 

For all components, the standard deviation of the angular velocity
$\underline{\Omega}_{b}^{(b)}$ first grows in time, until some saturation
is reached. The initial growth rate is largest for the roll component
(Fig. \ref{fig:FF_tu033_lambda077_angles}A), which presents large
fluctuations at $t=2T$ already. By this time, the pitch component
(B) has almost vanishing fluctuations and those in yaw (C) are significantly
smaller. The insects motion is damped by the viscous fluid, and thus
the magnitude of the angular velocity remains bounded.

Fig. \ref{fig:FF_tu033_lambda077_angles}D--F show the same quantities
as Fig. \ref{fig:FF_tu033_lambda077_angles}A--C for the case $\Lambda=0.32R$.
While the qualitative behavior is similar, the magnitude of both changes
in angular orientation and angular velocities of the body ($\underline{\Omega}_{b}^{(b)}$)
are significantly reduced. For example, $\gamma_{\mathrm{yaw}}=20^{\circ}$
in the $\Lambda=0.77R$ case is reduced to $2.5^{\circ}$. The fluctuations
in roll angular velocity grow fastest.

\begin{figure}[h]
\noindent \begin{centering}
\includegraphics[width=0.8\textwidth]{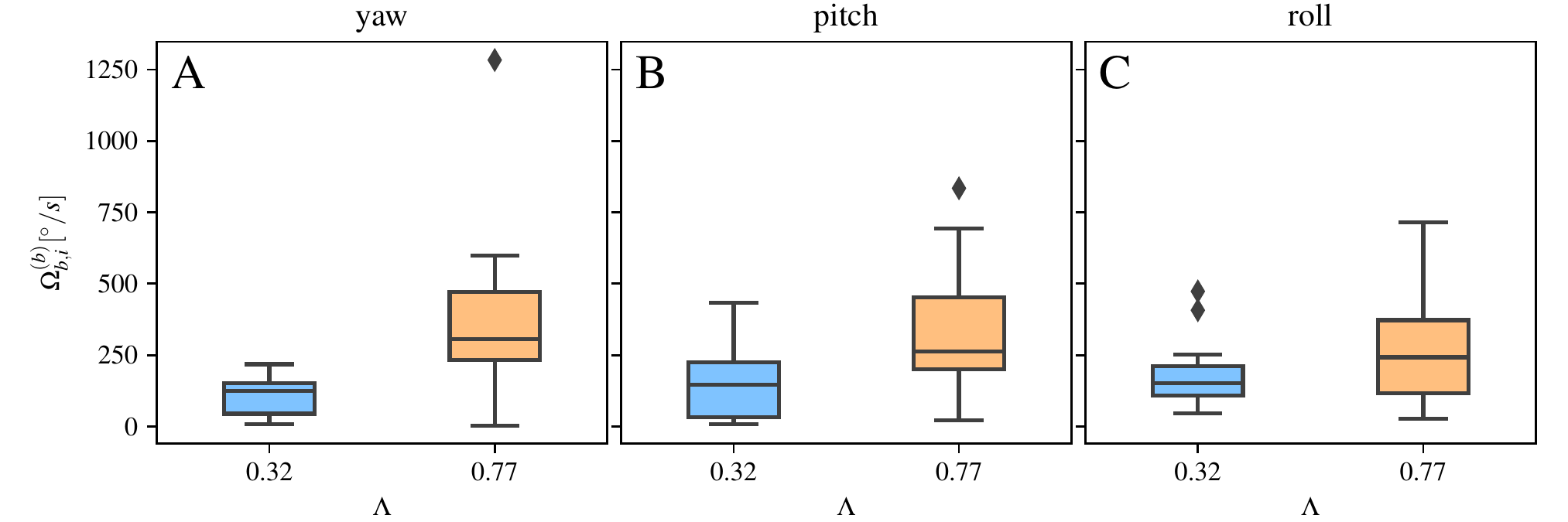}
\par\end{centering}
\caption{\label{fig:Free-flight-in-turbulent-lambda-ERRORBAR} Free flight
in turbulent inflow. Yaw (A), pitch (B) and roll (C) components of
the body angular velocity, averaged over the last computed cycle,
as a function of $\Lambda$. Data are represented by a box plot. In
the colored boxes, the line is the median of the data, and the limits
of the box is the upper and lower quartiles. The additional vertical
line is min/max values excluding outliers ($\diamondsuit$).}
\end{figure}

From the direct comparison of the two cases we can confirm the conclusions
from the tethered simulations also in the free flight case. The reduced
integral scale significantly reduces the impact of the flow on the
insect's attitude. Fig. \ref{fig:Free-flight-in-turbulent-lambda-ERRORBAR}
shows the magnitude of the different components of the angular velocity
and confirms that conclusion. Furthermore, as the 95\% confidence
intervals of the different directions overlap for both values of $\Lambda,$
again no direction with statistically significantly increased magnitude
can be observed. It appears thus from Fig. \ref{fig:FF_tu033_lambda077_angles}
that while the roll angular velocity grows fastest, its terminal value
is not significantly larger than the other two components, yaw and
pitch.

A key advantage of numerical work is that we can exclude any voluntary
motion that might be used, \emph{e.g.}, for distance estimation \citep{Ravi2016}.
However, at this point, we cannot give a quantitative estimate for
the limit of stable flight in turbulent conditions. The first uncertainty
concerns the degree of desired control. Experimental work \citep{Ravi2016}
suggests that bumblebees passively ride out small scale perturbations
and actively impose a long wavelength casting motion. It thus seems
that real animals are somewhat behaving between the two limiting cases
of tethered and free flight. This can also result from energetic considerations
as allowing for a certain amount of deviations may reduce overall
energetic cost.

The role of the reaction time delay appears to be the second crucial
factor for evaluating the stability. Our free flight data shows that
fluctuations in angular velocity grow fastest for the roll axis, which
is a consequence of the reduced moment of inertia. Figs. \ref{fig:FF_tu033_lambda077_angles}A
and C show that the roll component has reached its saturation at about
$t=2T$. Beyond this time, damping inhibits further growth, possibly
via the flapping counter torque (FCT) mechanism \citep{Hedrick2009,Hedrick2011}.
This does not imply any bound for changes in body angles, which continuously
grow in time. However, without the damping, the angular velocities
are expected to grow continuously, leading to much greater changes
in orientation.

Experimental work \citep{Vance2013} showed that honeybees (\emph{Apis
mellifera}) use angular velocities for roll, pitch and yaw of $3090,\,\,697,\,\,1874\,^{\text{\ensuremath{\circ}}}\mathrm{/s}$,
respectively, during the active recovery phase after being perturbed
with a strong wind gust. The magnitude of this angular velocity is
$3680\,^{\circ}\mathrm{/s}$, which is higher than the largest value
we find in our simulations (Fig. \ref{fig:FF_tu016-99}), and also
higher than the $2060\,^{\circ}/s$ which \citep{Vance2013} reports
during the passive phase directly after the perturbation. The associated
reaction time is stated as $3.5T<\tau_{\mathrm{react}}<6T$. Besides
differences in species (we are not aware of data available for bumblebees
in the literature), the study cannot directly be used to define a
threshold for the angular velocity beyond which the animals cannot
recover. In addition, flying in turbulence imposes continuous perturbations,
while \citep{Vance2013} studied the effect of a singular gust.

\section{Conclusions and perspectives\label{sec:Conclusions-and-perspectives}}

We numerically studied the impact of turbulence on a model insect,
using high-resolution numerical simulations on massively parallel
machines. Both tethered and free flight without control have been
considered, using a bumblebee model with rigid wings and prescribed
wing beat kinematics. The inflow condition ranged from laminar to
turbulent, and in the latter we varied the turbulence intensity as
well as the spectral distribution of the turbulent kinetic energy.
For the turbulent inflows, we performed ensemble-averaging to obtain
statistical estimates of forces, moments and power in tethered case
and body orientation and velocities in the free flight case.

In tethered flight, we have statistically estimated that the turbulent
inflow does not induce the detachment of the leading edge vortex.
This is true even in the strongest turbulence case and has already
been shown in our previous work \citep{Engels2015}. In addition to
the turbulence intensity, here we found the spectral distribution
of turbulent kinetic energy to be a significant parameter to be taken
into account. If the integral scale of the inflow is smaller than
the wing length, we found that statistically perturbations are reduced
for forces, moments and power, compared to turbulent inflow with larger
integral scale. We have demonstrated that the pressure field of the
turbulent perturbations is associated likewise with large scale variations
if the integral scale is large. The positive and negative pressure
perturbations have thus less chance of canceling out over the body,
which induces larger fluctuations.

Using free flight simulations we first checked that our model remains
stable for laminar inflow condition. In turbulent inflow, we confirmed
the finding from the tethered flight. We found that changes in body
orientation and angular velocity are highly sensitive to variations
in the turbulence spectrum. For constant turbulence intensity, a smaller
integral scale results in much smaller angular velocities and changes
in orientation. By modifying the turbulence intensity at fixed integral
scale, we showed how the angular velocities increase when the perturbations
become stronger. In all free flight simulations, we found the translation
of the insect to be small compared to its rotational motion.

Collectively, our findings suggest that the scales of turbulent motion
have a significant effect on the aerodynamics of flapping flight and
should hence be considered in future contributions on this topic.

In perspective, we plan to overcome the limitations of the current
study and specifically include the effects of both wing flexibility
and flight control. Moreover, since our results have been obtained
using a single species, namely a bumblebee, the generalization to
other insects is another important direction for future work. Finally,
we aim to replace the homogeneous isotropic turbulence, which is a
valuable starting point, by generic turbulent flows even more relevant
to insects, e.g., flower wakes.
\begin{acknowledgments}
Financial support from the Agence nationale de la recherche (ANR Grant No. 15-CE40-0019)
and Deutsche Forschungsgemeinschaft (T.E., J.S.: DFG Grant No. SE 824/26-1; F.L.: Grant No. LE
905/17-1), project AIFIT, is gratefully acknowledged. D.K. gratefully acknowledges financial sup-
port from the JSPS KAKENHI Grant No. JP18K13693. The authors were granted access to the HPC
resources of IDRIS under the Allocation No. 2018-91664 attributed by GENCI (Grand Équipement
National de Calcul Intensif). For this work, we were also granted access to the HPC resources
of Aix-Marseille Université financed by the project Equip@Meso (No. ANR-10-EQPX- 29-01).
T.E., K.S., M.F., F.L., and J.S. thankfully acknowledge financial support granted by the ministères
des Affaires étrangères et du développement international (MAEDI) et de l? Education national et l?enseignement supérieur, de la recherche et de l?innovation (MENESRI), and the Deutscher
Akademischer Austauschdienst (DAAD) within the French-German Procope project FIFIT.

\appendix
\end{acknowledgments}

\section{Appendix: Convergence to an infinitesimally thin flapping wing }

\begin{figure}[h]
\begin{centering}
\includegraphics[width=0.5\textwidth]{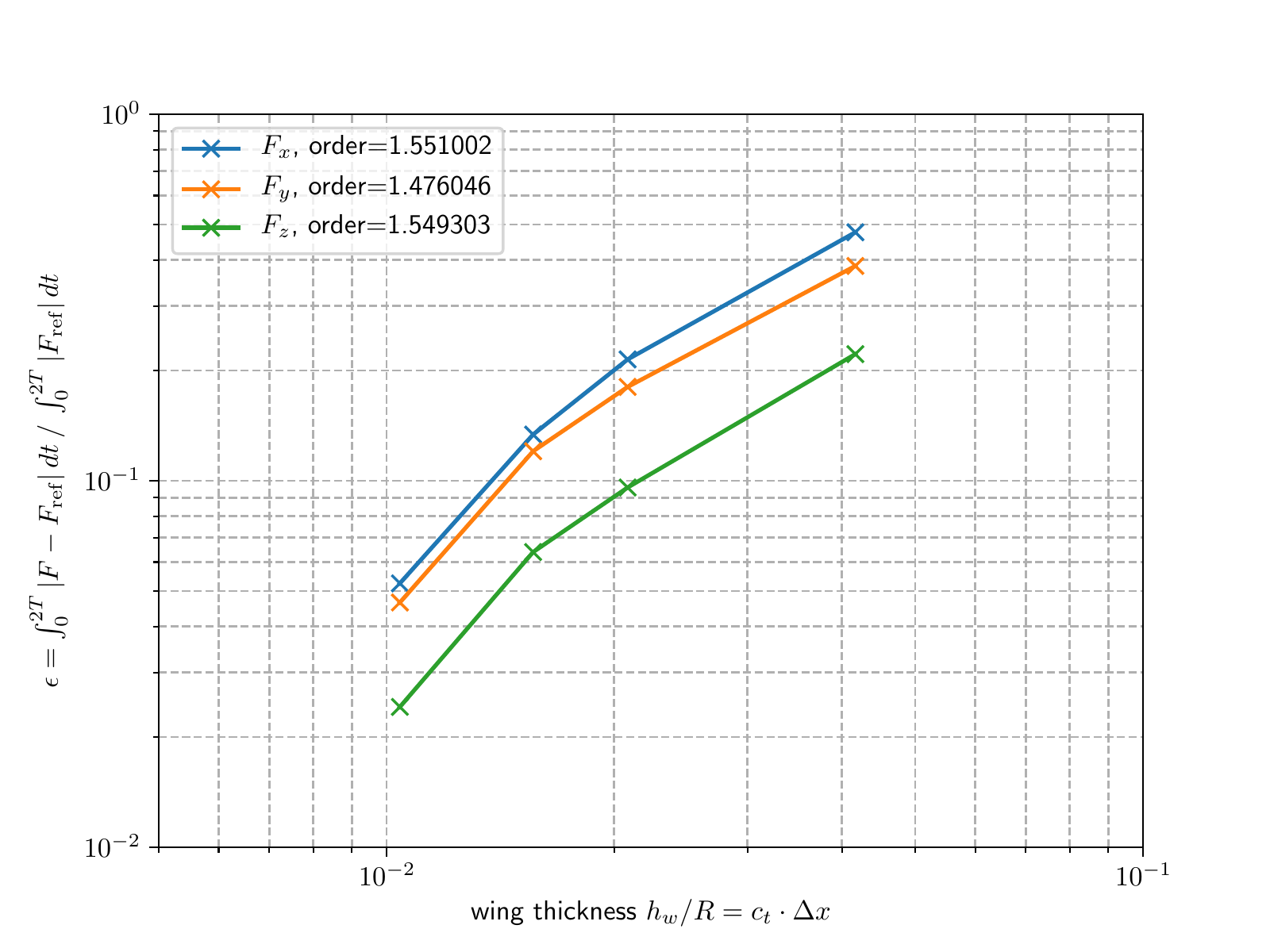}
\par\end{centering}
\caption{Convergence of forces in the wing thickness.\label{fig:Convergence-of-forces}}
\end{figure}

In this appendix, we study the convergence of our
numerical scheme in the limit of infinitesimally thin wings. We choose
the same wing geometry as in the rest of the article, with the same
kinematics, but simulate only one wing without the insects body. The
domain size is reduced to $2\times2\times2$ in order to be able to
reach high resolutions. The thickness of the wing is $h_{w}/R=c_{t}\Delta x$
where we set the constant $C_{t}=4$. As no reference solution is
available, we instead use the solution on the finest grid. As described
in \citep{Engels2015a}, the penalization parameter $C_{\eta}=(K_{\eta}\Delta x)^{2}/\nu$
is reduced with increasing resolution, in order to achieve optimal
results. The constant is $K_{\eta}=7.4\cdot10^{-2}$. We perform five
simulations with resolution $192^{3}$, $384^{3}$, $512^{3}$, $768^{3}$
and $1024^{3}$, with $h_{w}/R$ ranging from 4.2\% to 0.78\%. The
error is evaluated as 
\[
\epsilon=\int_{0}^{2T}(F_{i}-F_{\mathrm{ref},i})\mathrm{d}t/\int_{0}^{2T}F_{\mathrm{ref},i}\mathrm{d}t.
\]
Fig. \ref{fig:Convergence-of-forces} shows the resulting convergence.
For all components, we find qualitatively the same behavior and an
order of about 1.5. We can hence conclude that the penalization method
retains its accuracy also in the limit of thin flapping wings.

\bibliographystyle{plain}
\bibliography{bibliography}

\end{document}